\documentclass{article}
\usepackage{ijcai23}

\usepackage{times}
\usepackage{soul}
\usepackage{url}
\usepackage[hidelinks]{hyperref}
\usepackage[utf8]{inputenc}
\usepackage[small]{caption}
\usepackage{graphicx}
\usepackage{amsmath}
\usepackage{amsthm}
\usepackage{booktabs}
\usepackage{algorithm}
\usepackage{algorithmic}
\usepackage[switch]{lineno}
\usepackage{tikz}
\usetikzlibrary{trees}
\usepackage{enumitem}
\usepackage{amssymb}
\usepackage{balance}

\usepackage{multirow}
\usepackage{multicol}

\newcommand{\stitle}[1]{\vspace{1ex} \noindent{\bf #1}}

\title{Poisoning Attacks and Defenses in Recommender Systems: A Survey}

\author{
Zongwei Wang$^1$\and
Junliang Yu$^2$\and
Min Gao$^1$\footnote{Corresponding Author}\and
Wei Yuan$^2$\and
Guanhua Ye$^3$\and
Shazia Sadiq$^2$\and
Hongzhi Yin$^{2*}$
\affiliations
$^1$Chongqing University \and
$^2$The University of Queensland \and
$^3$Beijing University of Posts and Telecommunications
\emails
$^1$\{zongwei, gaomin\}@cqu.edu.cn \and
$^2$\{jl.yu@, w.yuan@, shazia@eecs., h.yin1@\}uq.edu.au \and
$^1$\{g.ye\}@bupt.edu.cn
}

\begin{document}

\maketitle

\begin{abstract}
Modern recommender systems (RS) have profoundly enhanced user experience across digital platforms, yet they face significant threats from poisoning attacks. These attacks, aimed at manipulating recommendation outputs for unethical gains, exploit vulnerabilities in RS through injecting malicious data or intervening model training. This survey presents a unique perspective by examining these threats through the lens of an attacker, offering fresh insights into their mechanics and impacts. Concretely, we detail a systematic pipeline that encompasses four stages of a poisoning attack: setting attack goals, assessing attacker capabilities, analyzing victim architecture, and implementing poisoning strategies. The pipeline not only aligns with various attack tactics but also serves as a comprehensive taxonomy to pinpoint focuses of distinct poisoning attacks. Correspondingly, we further classify defensive strategies into two main categories: poisoning data filtering and robust training from the defender's perspective. Finally, we highlight existing limitations and suggest innovative directions for further exploration in this field.
\end{abstract}

\section{Introduction}
Modern Recommender Systems (RS) have exhibited unprecedented capability in enhancing user experience by effectively matching the diverse preferences of individual users with an enormous number of available items \cite{42wang2023efficient,43cheng2016wide,c6wang2019minimax}. However, it has been increasingly recognized that these advanced RS show vulnerabilities to malicious activities, notably to Poisoning Attacks \cite{50gunes2014shilling,51si2020shilling,85zhang2020gcn}. Such poisoning attacks integrate deceptive or harmful data into the RS training datasets or influence the model gradient propagation in the training process, to compromise system functionality and manipulate outcome recommendations. 
For instance, e-commerce RS may inadvertently promote low-quality products when an attacker manipulates malicious user's interactions with these products, thereby compromising the system's integrity. Similarly, in the realm of news distribution, altering key phrases or visuals in news headlines or content can lead to the intentional dissemination of misinformation to specific user groups, potentially threatening societal stability. 
The presence of poisoning attacks undermines the original purpose of RS, transforming these platforms into tools that can be exploited for malicious intents.

Initial research into poisoning attacks in RS primarily explored heuristic-based methods, yielding a wide variety of well-developed hand-engineered attack models. In its infancy stages, the studies focused primarily on basic attack models such as the primal sample attack, random attack, average attack \cite{c1lam2004shilling}, and love/hate attack \cite{c2turk2019robustness}, which necessitated minimal understanding of the rating data. As insights into attacker tactics deepened, researchers uncovered more sophisticated attacks. For example, the bandwagon attack \cite{50gunes2014shilling} utilizes Zipf's law of user-item relationships to strategically select items that strengthen connections between target items and popular items. The relation attack \cite{48yu2017hybrid} goes beyond mere item manipulation, targeting the dynamics within social recommendation algorithms. These methods involve pre-defining a fixed strategy for generating and injecting malicious profiles into the system. Nevertheless, the static nature of these strategies often led to their predictability, rendering them detectable to defense measures once their patterns are deciphered~\cite{16wang2022gray}. 
Thus, there has been a paradigm shift towards more sophisticated poisoning attacks. 

Recent studies in this field have increasingly concentrated on tailoring attack techniques to accommodate both decentralized and centralized RS architectures. Different architectures introduce distinct vulnerabilities, requiring attackers to develop more complex and customized strategies to exploit specific weaknesses within these systems~\cite{c8wang2022fast,c9yuan2023federated,c23yang2020federated,c24long2023model}. Additionally, a thorough understanding of the attackers' capabilities within real-world scenarios is entailed
~\cite{16wang2022gray,08fan2021attacking,05christakopoulou2019adversarial}. In situations where the victim model or the details of the training dataset remain unknown, attackers are compelled to strategically allocate their economic resources to maintain stealth and effectiveness. These constraints represent substantial hurdles for attackers seeking to implement impactful strategies~\cite{07lin2020attacking,74zeng2023practical}. Moreover, the choice of poisoning strategy—whether to manipulate data or to modify model gradients—is crucial. Each approach targets specific vulnerabilities within the RS, underscoring the necessity for strategic planning on the part of attackers in selecting the most effective method to achieve their intended impact.

Gaining insights into the investigation of poisoning attacks, researchers have developed two primary defensive strategies for safeguarding RS: The first branch is to filter out poisoning information, which involves pinpointing and eliminating malicious elements within the system, such as counterfeit users exhibiting suspect behaviors or items infused with misleadingly negative information~\cite{d18xu2019detecting,d13wang2022detecting,d15hao2023detection}. This branch hinges on the application of supervised anomaly filtering, unsupervised clustering filtering, and confidence-based filtering strategies. The second strategy focuses on enhancing the robustness of recommendation models. This approach includes bolstering model training, for example, through mutual enhancement among multiple models or incorporating adaptive learning of confidence levels for each sample into the training process~\cite{a26wu2021fight,a31wang2021denoising,a10wang2022learning}. It also involves information enrichment or augmentation, such as introducing additional data (e.g., attributes, text, images) to enrich interaction data from existing data.

To safeguard RS against threats of poisoning attacks, it is imperative for researchers to comprehend how attackers fulfill their malicious activities. The purpose of this survey is to deepen the understanding of researchers and practitioners in this area by offering a systematic review of the literature on poisoning attacks against recommendation and their defense. In this survey, we detail a systematic pipeline that outlines four key stages critical in orchestrating poisoning attacks, which also serves as a comprehensive taxonomy to identify the focal points of various poisoning attacks. These stages include setting attack goals, understanding attacker capabilities, analyzing victim architecture, and implementing poisoning strategies.
On this basis, we delve into defense strategies against poisoning attacks from the defender's perspective, segmenting them into two phases: poisoning data filtering before malicious elements enter the system and robust training after the introduction of poisoning elements into the system.
In summary, our contributions are threefold: 

\begin{itemize}[leftmargin=*]
    \item We introduce a systematic pipeline that functions as a comprehensive taxonomy from the attacker's perspective, aimed at providing a fresh understanding of poisoning attacks against recommender systems.
    \item We encapsulate existing defense mechanisms and categorize them according to differences in their defensive stages. Additionally, we further break down each category of defenses from a technological standpoint.
    \item We shed light on the current challenges and promising future directions from both attack and defense perspectives. These critical insights will spur further innovation and exploration in secure RS.
\end{itemize}

\stitle{Connections to Existing Surveys:}
While previous surveys have investigated trustworthy~\cite{52fan2022comprehensive} and robust~\cite{53anelli2021adversarial} RS, the specific vulnerabilities arising from poisoning attacks have not been thoroughly explored. There are existing studies summarizing heuristic-based~\cite{51si2020shilling} and adversarial learning-based poisoning attacks~\cite{54deldjoo2021survey}, yet they fail to fully cover the broadening scope of research in the latest poisoning attacks, such as those involving contrastive learning-based~\cite{82wang2023poisoning,30wang2023poisoning} and large language model-based recommendations~\cite{c4zhang2024stealthy,27wu2023attacking}. Additionally, a separate study~\cite{88rezaimehr2021survey} focuses solely on examining detection techniques in collaborative filtering-based RS, neglecting wider discussions on defense strategies during the model training phase. By conducting an extensive review of the existing literature, we aim to provide a detailed understanding of poisoning attacks and their defense mechanisms, guiding both academic and industrial efforts toward enhancing the security of RS.

\stitle{Paper Collection:} 
In this survey, we conducted a thorough review of 48 latest high-quality research papers focusing on poisoning attacks and 46 research papers focusing on defense. Our literature search was conducted primarily using DBLP and Google Scholar, with the keywords ``attack/poisoning + recommendation/recommender'' for poisoning attack research and keywords `` defense/detection/robust + recommendation/recommender''. We then traversed the citations of the identified papers and included relevant studies. To capture the most influential and cutting-edge work, we consistently monitored leading conferences and journals, including IJCAI, ICDE, CIKM, KDD, WWW, SIGIR, WSDM, TKDE, etc. Additionally, to encompass emerging trends and innovative concepts, we included preprints from arXiv that presented novel ideas to this field. It should be noted that the overarching aim of this survey is to provide an updated landscape of this area. Consequently, this study delivers only a concise summary of those papers published before 2018 in this field instead of a detailed analysis.


\stitle{Survey Structure:} The structure of this survey is organized as follows: Section 2 delineates the conceptual framework and formalization of the victim model, alongside the mechanisms of poisoning attacks and defensive strategies within the RS domain. Section 3 categorizes poisoning attacks into four distinct types based on the motivations outlined in the original papers and delineates defense strategies into two primary categories according to their protective objectives. Section 4 delves into the exploration of poisoning attack methods, following the established taxonomy. Section 5 offers an in-depth examination of defensive approaches. 
In Section 6, we have summarized some commonly used evaluation metrics about poisoning attacks and defenses in recommendation domains.
Section 7 highlights current limitations and proposes potential avenues for future research. The survey concludes with Section 8.

\section{Preliminary}
\subsection{Victim Recommender System}
Let $\mathcal{U}$ and $\mathcal{I}$ represent the sets of users and items in a RS, with $\mathcal{D}$ indicating the original user/item dataset that includes interaction details (such as clicks, likes, and collections) and features (like text and images).
A recommender function $f_{\theta}$ is trained on the dataset $\mathcal{D}$ to learn a low-dimensional representation, which is used to predict items from $\mathcal{I}$ that might interest a user $u \in \mathcal{U}$. Formally:
\begin{equation}
\begin{aligned}
\mathbf{\Theta}^{*}=\arg\min\limits_{\mathbf{\Theta}}\mathcal{L}_{rec}(f_{\theta}(\mathcal{D})),
\end{aligned}
\label{bi-level optimization}
\end{equation}
where $\mathbf{\Theta}^{*}$ represents the optimal model parameters, including user and item representations, and $\mathcal{L}_{rec}$ is the recommendation objective function such as Cross-Entropy~\cite{c7de2005tutorial}.

In the context of poisoning attacks, the recommendation model $f_{\theta}$ is the system that the attacker intends to manipulate, referred to as the ``victim model''.

\subsection{Attacks on Recommender Systems}
Attackers can control a subset of malicious users $\mathcal{U}_{M}$, who engage in strategic interactions with various items, occasionally altering some item features. The resulting malicious dataset $\mathcal{D}_{M}$ is incorporated into the training data of the RS. The objective of these attacks is to distort the learning process of recommendation models to achieve specific goals. In this context, we define the poisoning attack task as a bi-level optimization problem:
\begin{equation}
\begin{aligned}
&\mathcal{D}_{M}=\arg\max\limits_{\mathcal{D}_{M}}\mathcal{L}_{attack}(\mathcal{D}, \mathcal{D}_{M}, \mathbf{\Theta}^{*}), \\
\mathrm{s.t.}&, \quad  \mathbf{\Theta}^{*}=\arg\min\limits_{\mathbf{\Theta}}\mathcal{L}_{rec}(f_{\theta}(\mathcal{D},\mathcal{D}_{M})).
\end{aligned}
\label{bi-level optimization}
\end{equation}
Here, $\mathcal{L}_{attack}$ represents the poisoning attack loss, which measures the attack's effectiveness, typically considering metrics such as the decline in recommendation accuracy and the disruption of the ranking of targeted items. The malicious data $\mathcal{D}_{M}$ is generated through alternating inner and outer optimization steps. The inner optimization involves training the victim model using both the original and malicious data. The outer optimization refines the malicious data to fulfill the attack's objectives. It's important to note that $f_{\theta}$ can be a surrogate model, as attackers may not have access to the actual victim model.

\subsection{Defense against Poisoning Attacks}
To mitigate the negative impacts of poisoning attacks on RS, the defenders of these systems adopt measures to filter out malicious elements from the system before model training. Alternatively, they may enhance the recommendation system's resilience to attacks by introducing robust model training strategies. We summarize this defense process with the following formulation:
\begin{equation}
\begin{aligned}
&\mathbf{\Theta}^{*}=\arg\min\limits_{\mathbf{\Theta}}\mathcal{L}_{rec}(\mathcal{D}',\mathbf{X}), \\
\mathrm{s.t.}, \quad & \mathcal{D}' = f_{filter} (\mathcal{D}), \quad \mathbf{X}= f_{robust}(\mathcal{D}),
\end{aligned}
\label{bi-level optimization}
\end{equation}
where $f_{filter}$ represents the malicious information filtering mechanism employed by the defender, where $\mathcal{D}'$ is the high-quality training data obtained after filtering. $f_{robust}$ denotes the defender's robustness enhancement strategy, resulting in $\mathbf{X}$, which consists of augmented information (e.g., enhanced representations and data confidence levels). Under these two defense strategies, the recommendation model leverages high-quality data and augmented information for training, thereby enhancing its resilience against poisoning attacks.

\section{Poisoning Attack and Defense Taxonomies}

\subsection{Taxonomy of Poisoning Attacks against Recommendation}
\begin{figure}[t]
\centering
\includegraphics[width=0.48\textwidth]{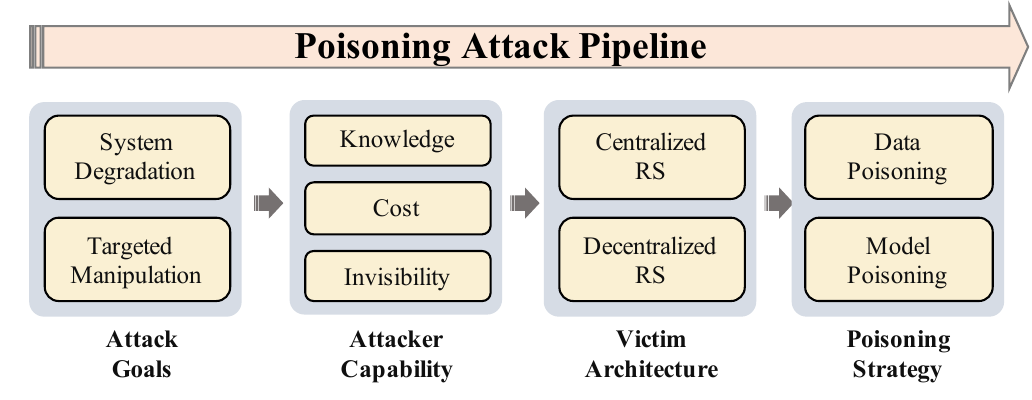}
\caption{The taxonomy of poisoning attacks against RS.}
\label{Taxonomy}
\end{figure}

To deepen our understanding of poisoning attacks, it is crucial to think about the fundamental stages of an attack from the perspective of attackers. Following this approach, a systematic pipeline is developed, as illustrated in Figure \ref{Taxonomy}. This pipeline is dissected into four distinct stages, collectively outlining the strategic progression of an attack. Initially, the attackers must determine the precise goal of the attack, pinpointing what they aim to disrupt or manipulate within the victim system. Next, they thoroughly assess their available capabilities, considering their knowledge and budgets to sketch potential methods of intrusion that remain invisible to safeguard systems. Subsequently, the attackers conduct a detailed analysis of the RS's architecture if known, scrutinizing it to uncover any specific vulnerabilities of which can be taken advantage. Ultimately, armed with this intelligence, the attacker strategically devises a tailored attack plan, selecting methods and tactics that are most likely to succeed in achieving their defined objectives. As the pipeline is fundamental to any type of poisoning attacks, the uncertainties and variations, such as the attacker's capabilities and the architecture of the victim system, can provide a fertile ground for various attack strategies. Regardless of the technical specifics, the pipeline can encapsulate these attacks and categorize them based on their focal points, serving effectively as a comprehensive taxonomy. In the subsequent sections, we will navigate the readers through each stage within this pipeline with detailed introductions.

\subsubsection{Setting Attack Goals}
\begin{figure}[h]
\centering
\includegraphics[width=0.45\textwidth]{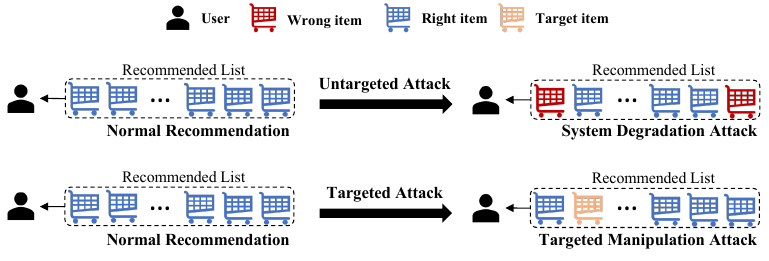}
\caption{Poisoning attacks for system degradation and targeted manipulation.}
\label{Goal}
\end{figure}

Through the lens of an attacker, the objectives of poisoning attacks are meticulously aligned with their strategic intent to manipulate or impair RS. The first category, known as \textbf{system degradation attacks} (a.k.a untargeted attacks), is crafted with the broad objective of undermining the entire system~\cite{37wu2023influence,20wu2022fedattack}. The attacker’s goal here is to erode the quality of all recommendations, thereby diminishing the overall user experience and potentially precipitating significant financial losses for the service provider. Such attacks are often orchestrated by competitors of the recommendation platform. In contrast, \textbf{targeted manipulation attacks} (a.k.a targeted attacks) demonstrate a more specific approach~\cite{09song2020poisonrec,21chen2022knowledge}. These attacks are elaborately crafted to either promote or demote specific items within distinct user groups or across all users. Typically, these are the maneuvers of individual product vendors aiming to manipulate market perception favorably. Figure \ref{Goal} delineates the stark contrasts between the broad-reaching impact of system degradation attacks and the precision of targeted manipulation attacks.

\subsubsection{Assessing Attacker's Capability}
\begin{figure}[h]
\centering
\includegraphics[width=0.45\textwidth]{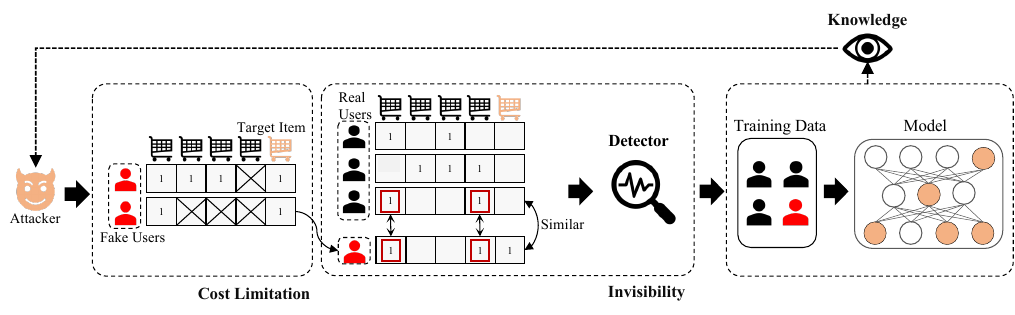}
\caption{Poisoning attacks constrained by prior knowledge, cost limitation and invisibility.}
\label{capibility}
\vspace{-10pt}
\end{figure}

To execute a successful poisoning attack on real-world RS, the attacker must take into account specific constraints, notably \textbf{knowledge}, \textbf{cost}, and \textbf{invisibility}. Figure \ref{capibility} illustrates the typical attack process, highlighting these considerations. 
 Attackers typically view RS as complex 'black boxes,' with limited transparency regarding the data used for training and the specifics of the underlying algorithms. Gaining an in-depth understanding of these components is crucial, as it not only informs the design of the attack but also maximizes the chances of its success by exploiting system vulnerabilities~\cite{16wang2022gray,32lin2022shilling}.
Financial considerations are integral to the attacker’s strategy. The insertion of poisoned data into the system requires a cost-effective analysis to ensure economic viability. Each act of data manipulation incurs inherent costs, from acquiring the necessary resources to executing the attack discreetly. Therefore, the attacker must meticulously plan the scale and frequency of these operations to stay within budget while still achieving the desired disruptive effect~\cite{08fan2021attacking,74zeng2023practical}.
Furthermore, the element of stealth is essential in avoiding detection. Poisoned data must be intricately crafted to blend seamlessly with legitimate data, making it challenging for the typical defense mechanisms within RS to detect the anomalies. This requires a sophisticated understanding of both the detection algorithms and the data patterns they are programmed to flag as suspicious. The attacker’s ability to maintain this invisibility not only prolongs the lifespan of the attack but also enhances its impact, as the longer the malicious data remains undetected, the more damage it can inflict. By meticulously balancing these factors, the attacker can orchestrate a potent and covert operation that significantly compromises the integrity of the RS~\cite{05christakopoulou2019adversarial,07lin2020attacking}.

\subsubsection{Analyzing Victim Architectures}
\begin{figure}[h]
\centering
\includegraphics[width=0.45\textwidth]{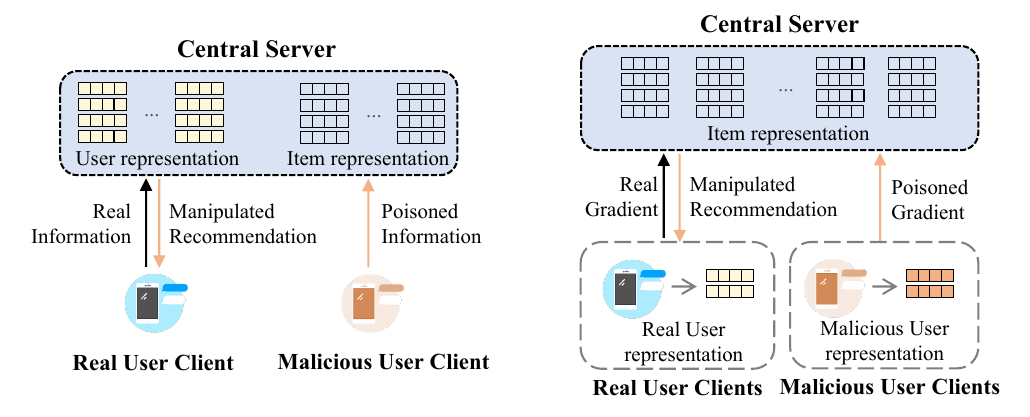}
\caption{Poisoning attacks against centralized and decentralized scenario. The left side of the illustration represents centralized RS, whereas the right side depicts decentralized RS.}
\label{Scenario}
\end{figure}

Before carrying out attacks, the attacker needs to analyze the architecture of the potential victim, a critical factor that shapes their approaches. These architectures are generally categorized into \textbf{centralized} and \textbf{decentralized} recommendations~\cite{c8wang2022fast,c9yuan2023federated,c23yang2020federated,c24long2023model}, as illustrated in Figure \ref{Scenario}.  In centralized RS, the consolidation of user and item data on a central server offers a lucrative target for attackers, who can disrupt the recommendation process across the entire user base by infiltrating this single hub. In contrast, decentralized RS (e.g., federated RS) distribute the training of the global model across numerous user devices, each retaining its sensitive data locally. This setup presents a more complex landscape for an attacker. Instead of a single point of attack, multiple nodes become potential targets, each requiring a unique strategy to introduce poisoned data effectively. The decentralized nature complicates the attacker’s task but also provides multiple avenues for discreet infiltration, making the detection and mitigation of such attacks more challenging. For an attacker, understanding these structural differences is vital. Centralized systems might be breached with a one-time, high-impact intervention, while decentralized systems might necessitate a stealthier, more sustained engagement. This nuanced understanding allows the attacker to craft strategies that exploit specific vulnerabilities related to each architecture type.

\subsubsection{Implementing Poisoning Strategies}
\begin{figure}[h]
\centering
\includegraphics[width=0.45\textwidth]{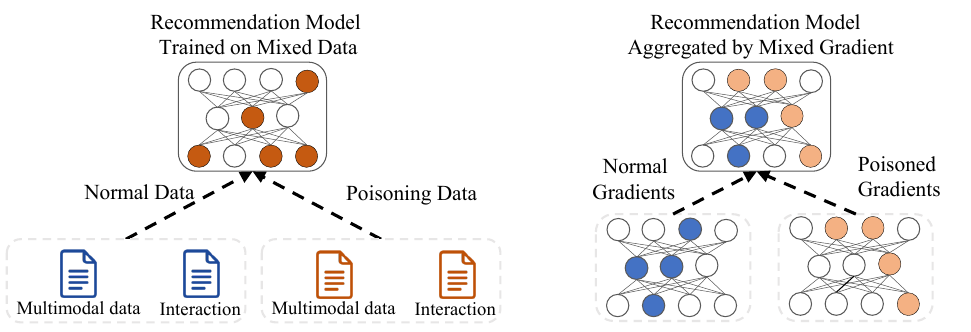}
\caption{Poisoning attacks divided by data poisoning and model poisoning.  The left is data poisoning, while the right is model poisoning.}
\label{ways}
\end{figure}

The architectural diversity of RS offers a playground for attackers, who can choose between two principal methods of infiltration: \textbf{data poisoning} and \textbf{model poisoning}, as depicted in Figure \ref{ways}. In the strategy of data poisoning, an attacker inserts malicious interaction or multimodal data into the RS. This method involves contamination of the training dataset, where the injected data blends imperceptibly with legitimate data, subtly degrading the training process and skewing the model’s output in favor of the attacker’s goals~\cite{01li2016data,02fang2018poisoning}. Model poisoning, in contrast, requires deeper, more technical manipulation. Here, the attacker targets the learning process itself by perturbing the model gradients. This involves the sophisticated generation and strategic upload of poisoned gradients during the aggregation phase, where information is pooled from multiple user devices~\cite{11zhang2022pipattack,19rong2022fedrecattack}. This tactic not only distorts the model's learning trajectory but also evades detection by embedding within the collaborative learning process, making it a particularly stealthy and effective method of attack.

\subsection{Taxonomy of Defense against Poisoning Attacks}
\begin{figure}[h]
\centering
\includegraphics[width=0.48\textwidth]{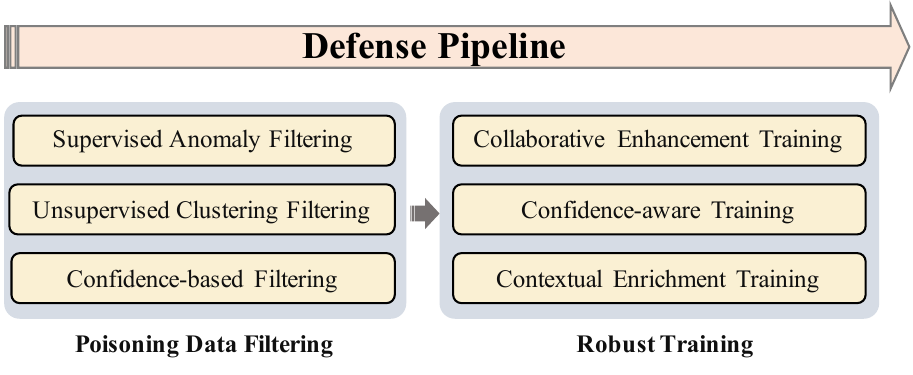}
\caption{The taxonomy of defense against poisoning attacks in RS.}
\label{Taxonomydefense}
\end{figure}

From the defender's perspective, crafting effective defenses against poisoning attacks involves a strategic and systematic categorization into two primary lines of defense, each corresponding to different stages of intervention, as outlined in Figure \ref{Taxonomydefense}. Initially, defenders adopt preemptive \textbf{poisoning data filtering} mechanisms. This strategy is grounded in the principle of `defense in depth', where the aim is to identify and isolate malicious elements before they can integrate into the system. By filtering out these potentially harmful inputs at an early stage, the majority of the threat is neutralized, thereby preserving the integrity and reliability of the recommender system. However, recognizing that preemptive measures may not catch all adversarial inputs, defenders also deploy \textbf{robust training} algorithms during the model training phase. This line of defense is designed to be resilient and equipped to detect and mitigate the influence of any malicious elements that might have eluded initial screening. These algorithms are often enhanced with machine learning techniques (e.g., adversarial training \cite{53anelli2021adversarial}) that adaptively learn to distinguish between benign and malicious patterns, thereby refining their defensive capabilities over time. Together, these categories form a holistic defense framework that is both descriptive and prescriptive, guiding defenders through a layered and adaptive security strategy. By framing these strategies within a taxonomy, defenders gain a clearer understanding of how and when to apply specific techniques, ensuring a comprehensive coverage that adapts to and addresses the spectrum of threats faced by recommender systems.

\subsubsection{Poisoning Data Filtering for Defense}
\begin{figure}[h]
\centering
\includegraphics[width=0.45\textwidth]{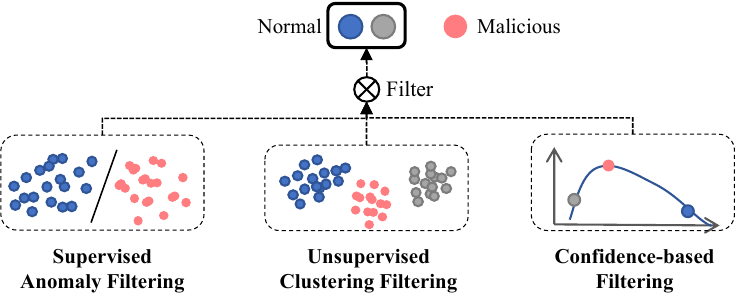}
\caption{The illustration of supervised anomaly filtering, unsupervised clustering filtering, and confidence-based filtering against poisoning attacks.}
\label{filter}
\end{figure}

Poisoning data filtering is a crucial defensive strategy where defenders identify and remove elements such as fake users exhibiting suspicious behaviors, items infused with negative information, or unusual connections between users and items before the data is utilized within the system. We categorize this filtering defense into three distinct types: \textbf{supervised anomaly filtering}, \textbf{unsupervised clustering filtering}, and \textbf{confidence-based filtering}. Figure \ref{filter} illustrates the distinctions among these categories.
Supervised anomaly filtering leverages labeled historical data to train models that can discern and isolate malicious entries based on learned patterns. This kind of approaches effectively create a boundary in the vector space that distinguishes between normal and malicious information, thereby enabling the system to filter out harmful data proactively\cite{d13wang2022detecting,d2hsiao2022unsupervised}.
Unsupervised clustering filtering is applied in scenarios lacking labeled data. It involves segregating data into various clusters without prior knowledge of their properties. This kind of method identify and filters out anomalies by detecting clusters that are significantly different in size or composition from the majority, isolating potentially malicious information based on its relational distance from normative clusters~\cite{d11cai2019bs,d17cai2019detecting}.
Lastly, confidence-based filtering employs probabilistic or statistical models to assess and score each piece of data according to its likelihood of being malicious. This technique filters out data points with high anomaly scores, effectively removing entries most likely to compromise the system's integrity~\cite{d3liu2020recommending,d1li2021large}.

\vspace{-10pt}
\subsubsection{Robust Training for Defense}
\begin{figure}[h]
\centering
\includegraphics[width=0.45\textwidth]{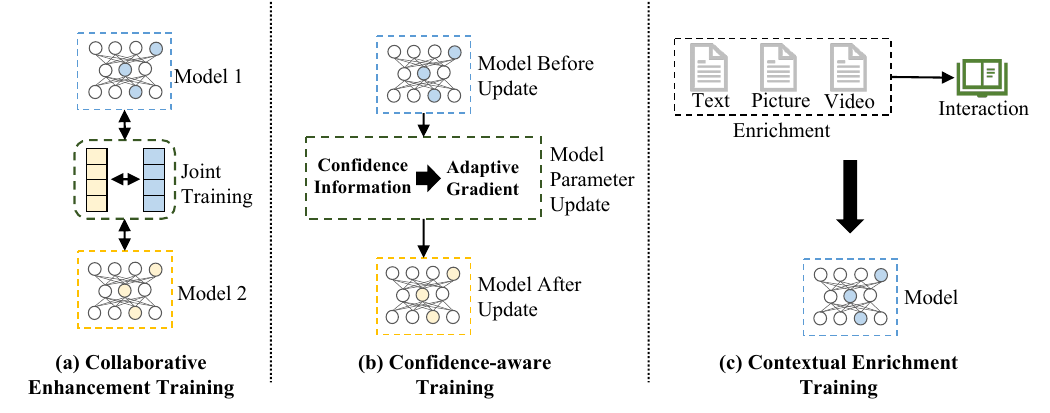}
\caption{The illustration of collaborative enhancement training, confidence-aware training, and contextual enrichment training.}
\label{robust training}
\end{figure}

Robust training employs strategies that inherently enhance the robustness of RS, allowing them to resist noise and attacks naturally. These strategies can be systematically categorized into three principal streams: \textbf{collaborative enhancement training}, \textbf{confidence-aware training}, and \textbf{contextual enrichment training}. Figure \ref{robust training} illustrates the distinctions among these strategies.
Collaborative enhancement training capitalizes on the synergy of multiple models working together, leveraging adversarial or collaborative methods to enhance each model’s resistance to attacks. This method not only strengthens individual model robustness but also fosters a cooperative defense mechanism across the system, enhancing overall resilience~\cite{a26wu2021fight,a13wang2023robust}.
Confidence-aware training, on the other hand, adaptively learns the malicious confidence level of each sample. By doing so, it prioritizes learning from reliable data and minimizes the impact of potentially poisoned inputs, thereby preserving the integrity and accuracy of the model’s outputs~\cite{a31wang2021denoising}. 
The last stream is dedicated to the integration of supplementary data signals. Contextual enrichment training expands the training dataset with a wide variety of additional signals such as user attributes, textual content, images, and social relations. This diversity in data sources enriches the model’s contextual understanding, enabling it to make more informed and nuanced predictions~\cite{a22bian2021denoising}.

\section{Poisoning Attacks against Recommender Systems}
Poisoning attacks are classified into four principal categories, each corresponding to a specific stage in the poisoning attack pipeline: attacks driven by malicious goals, attacks constrained by practical capabilities, attacks tailored to the victim's architecture, and attacks distinguished by poisoning strategies. Our taxonomy, which adopts a pipeline perspective, emphasizes the interconnected nature of these categories rather than treating them as isolated.  For instance, the \textit{GSPAttack}~\cite{33nguyen2022poisoning} aims to execute a targeted attack by promoting a specific item to more recommendation lists. It specifies that the model can only incorporate a limited number of malicious users and targets a graph-based recommendation model. The attack is then executed by manipulating the interaction data between users and items. To avoid redundancy, in subsequent sections, we will select and discuss typical attack methods from each category, providing a detailed analysis of how these strategies operate within their respective contexts. Furthermore, the relationship between each attack method and the four identified categories is summarized in Table \ref{summary of poisoning attacks}. 

\subsection{Attacks Driven by Malicious Goals}
Aligned with the strategic objectives of poisoning attacks, the attack goals are categorized into \textbf{system degradation attacks} and \textbf{targeted manipulation attacks}. The former type aims to compromise the overall system integrity, whereas the latter is engineered to either elevate or suppress specific items within particular user groups or among all users. Additionally, more sophisticated poisoning attacks are developed under the category of \textbf{hybrid-goals attacks}.
\subsubsection{System Degradation}
In untargeted system degradation attacks, one common strategy is to select specific users or items and then predominantly poison the RS through these targeted samples. For instance, \textit{Infmix}~\cite{37wu2023influence,38wu2021fight} concentrates on identifying ``influential users" among malicious users who can exert a more significant impact on the system. The approach prioritizes the creation of interactions for these ``influential users" to amplify the attack's impact. Analogously, \textit{FedAttack}~\cite{20wu2022fedattack} enhances the attack effect from the perspective of items by utilizing globally ``hardest items" to disrupt model training. It selects items most and least relevant to user embeddings as the hardest negative and positive samples, respectively. Manipulating the interaction records of these hardest items significantly influences the system's outcomes.

In addition, another strategy stresses disrupting the orderly distribution of representations, thereby disturbing the recommendation outcomes. \textit{ClusterAttack}~\cite{23yu2023untargeted} controls poisonous profiles to cluster item embeddings into dense groups. This clustering causes the recommender to generate similar scores for items within the same cluster, thereby disrupting the ranking order. On the other hand,  \textit{UA-FedRec}~\cite{26yi2023ua} targets the user and item modeling processes, modifying the representations to increase the distance between similar samples and decrease the distance between dissimilar ones. This modification is contrary to the objectives of the recommendation model, thereby disrupting the overall recommendation process.

\subsubsection{Targeted Manipulation}
In a typical targeted attack scenario, attackers aim to significantly influence the placement of target items in the top-K recommendation lists for as many users as possible. To achieve this, attackers have designed sophisticated attack functions to optimize the attack model. For example, \textit{PoisonRec}\cite{09song2020poisonrec} focuses on maximizing the frequency of targeted items appearing in top-K recommendation lists by manipulating interaction data to boost item visibility across user profiles. Alternatively, \textit{A-hum}~\cite{22rong2022poisoning} aims to directly maximize the predicted probability of each target item being recommended to users by influencing the model to assign higher relevance scores to these items during the recommendation process. Similarly, \textit{RAPU}~\cite{12zhang2021data} seeks to ensure that the predicted probabilities of the target item being recommended exceed those of other items by adjusting the training data to favor the target item, ensuring it consistently ranks higher in recommendation predictions. In addition to directly promoting or demoting specific items, some poisoning attack approaches design unique and specialized objectives to achieve similar outcomes. A notable example is \textit{GTA}~\cite{40wang2023revisiting},  which first evaluates user intent by predicting the most favored item for each user. It then adjusts the target item to closely mimic these preferred items. This process alters the target item's attributes to match those of user-preferred items, subtly affecting the recommendation outcomes.

Another variant of of targeted attacks in poisoning involves the deliberate promotion of certain items to a specific group of users. This strategy ensures that these chosen users encounter the items selected by the attacker. For example, \textit{AutoAttack}~\cite{73guo2023targeted} considers both the target item and user perspectives and it crafts malicious user profiles that closely mimic the attributes and interactions of the targeted user group. This strategy enables an effective influence on the designated group while concurrently minimizing unintended effects on other users. Analogously, \textit{UBA}~\cite{c5wang2024uplift} proposes that each item has a distinct target audience and that the challenge of attacking various users differs. A common approach targeting all users can lead to the wasteful expenditure of a fake user budget and a decline in attack efficacy. Therefore, \textit{UBA} introduces an attack method that considers both performance and budget from a causal perspective, aiming for efficient utilization of resources while maximizing impact to promote targeted items to specific user groups. 

\begin{table*}[h]
\caption{A summary of the poisoning attack methods against recommendation.} 
\centering
\resizebox{0.75\textwidth}{!}{
\begin{tabular}{@{}c|cc|ccc|cc|cc@{}}
\toprule
{\textbf{Model}} &
  \multicolumn{2}{c|}{\textbf{Attack Goal}} &
  \multicolumn{3}{c|}{\textbf{Attacker Capability}} &
  \multicolumn{2}{c|}{\textbf{Victim Architecture}} &
  \multicolumn{2}{c}{\textbf{Poisoning Way}} \\ \cmidrule(l){2-10} 
 &
  \multicolumn{1}{c|}{Untargeted} &
  Targeted &
  \multicolumn{1}{c|}{Knowledge} &
  \multicolumn{1}{c|}{Cost} &
  Invisibility &
  \multicolumn{1}{c|}{Centralized} &
  Decentralized &
  \multicolumn{1}{c|}{Data Poisoning} &
  Model Poisoning \\ \midrule
TNA~\cite{34fang2020influence} &
  \multicolumn{1}{c|}{} &
  $\checkmark$ &
  \multicolumn{1}{c|}{$\checkmark$} &
  \multicolumn{1}{c|}{} &
   &
  \multicolumn{1}{c|}{$\checkmark$} &
   &
  \multicolumn{1}{c|}{$\checkmark$} &
   \\ \midrule
UNAttack~\cite{31chen2021data} &
  \multicolumn{1}{c|}{} &
  $\checkmark$ &
  \multicolumn{1}{c|}{$\checkmark$} &
  \multicolumn{1}{c|}{} &
   &
  \multicolumn{1}{c|}{$\checkmark$} &
   &
  \multicolumn{1}{c|}{$\checkmark$} &
   \\ \midrule
DLAttack~\cite{18huang2021data} &
  \multicolumn{1}{c|}{} &
  $\checkmark$ &
  \multicolumn{1}{c|}{$\checkmark$} &
  \multicolumn{1}{c|}{} &
   &
  \multicolumn{1}{c|}{$\checkmark$} &
   &
  \multicolumn{1}{c|}{$\checkmark$} &
   \\ \midrule
NCFAttack~\cite{25zhang2020towards} &
  \multicolumn{1}{c|}{} &
  $\checkmark$ &
  \multicolumn{1}{c|}{$\checkmark$} &
  \multicolumn{1}{c|}{$\checkmark$} &
   &
  \multicolumn{1}{c|}{$\checkmark$} &
   &
  \multicolumn{1}{c|}{$\checkmark$} &
   \\ \midrule
GOAT~\cite{15wu2021ready} &
  \multicolumn{1}{c|}{} &
  $\checkmark$ &
  \multicolumn{1}{c|}{$\checkmark$} &
  \multicolumn{1}{c|}{} &
  $\checkmark$ &
  \multicolumn{1}{c|}{$\checkmark$} &
   &
  \multicolumn{1}{c|}{$\checkmark$} &
   \\ \midrule
GSPAttack~\cite{33nguyen2022poisoning} &
  \multicolumn{1}{c|}{} &
  $\checkmark$ &
  \multicolumn{1}{c|}{$\checkmark$} &
  \multicolumn{1}{c|}{$\checkmark$} &
   &
  \multicolumn{1}{c|}{$\checkmark$} &
   &
  \multicolumn{1}{c|}{$\checkmark$} &
   \\ \midrule
SSLAttack~\cite{30wang2023poisoning} &
  \multicolumn{1}{c|}{} &
  $\checkmark$ &
  \multicolumn{1}{c|}{$\checkmark$} &
  \multicolumn{1}{c|}{$\checkmark$} &
   &
  \multicolumn{1}{c|}{$\checkmark$} &
   &
  \multicolumn{1}{c|}{$\checkmark$} &
   \\ \midrule
PromptAttack~\cite{27wu2023attacking} &
  \multicolumn{1}{c|}{} &
  $\checkmark$ &
  \multicolumn{1}{c|}{$\checkmark$} &
  \multicolumn{1}{c|}{} &
   &
  \multicolumn{1}{c|}{$\checkmark$} &
   &
  \multicolumn{1}{c|}{$\checkmark$} &
   \\ \midrule
CLeaR~\cite{82wang2023poisoning} &
  \multicolumn{1}{c|}{} &
  $\checkmark$ &
  \multicolumn{1}{c|}{$\checkmark$} &
  \multicolumn{1}{c|}{$\checkmark$} &
   &
  \multicolumn{1}{c|}{$\checkmark$} &
   &
  \multicolumn{1}{c|}{$\checkmark$} &
   \\ \midrule
H-CARS~\cite{28chen2023dark} &
  \multicolumn{1}{c|}{} &
  $\checkmark$ &
  \multicolumn{1}{c|}{$\checkmark$} &
  \multicolumn{1}{c|}{} &
   &
  \multicolumn{1}{c|}{$\checkmark$} &
   &
  \multicolumn{1}{c|}{$\checkmark$} &
   \\ \midrule
FedRecAttack~\cite{19rong2022fedrecattack} &
  \multicolumn{1}{c|}{} &
  $\checkmark$ &
  \multicolumn{1}{c|}{$\checkmark$} &
  \multicolumn{1}{c|}{$\checkmark$} &
  $\checkmark$ &
  \multicolumn{1}{c|}{} &
  $\checkmark$ &
  \multicolumn{1}{c|}{} &
  $\checkmark$ \\ \midrule
PSMU~\cite{29yuan2023manipulating} &
  \multicolumn{1}{c|}{} &
  $\checkmark$ &
  \multicolumn{1}{c|}{$\checkmark$} &
  \multicolumn{1}{c|}{} &
   &
  \multicolumn{1}{c|}{} &
  $\checkmark$ &
  \multicolumn{1}{c|}{} &
  $\checkmark$ \\ \midrule
A-hum~\cite{22rong2022poisoning}&
  \multicolumn{1}{c|}{} &
  $\checkmark$ &
  \multicolumn{1}{c|}{$\checkmark$} &
  \multicolumn{1}{c|}{} &
   &
  \multicolumn{1}{c|}{} &
  $\checkmark$ &
  \multicolumn{1}{c|}{} &
  $\checkmark$ \\ \midrule
PipAttack~\cite{11zhang2022pipattack} &
  \multicolumn{1}{c|}{} &
  $\checkmark$ &
  \multicolumn{1}{c|}{$\checkmark$} &
  \multicolumn{1}{c|}{} &
  $\checkmark$ &
  \multicolumn{1}{c|}{} &
  $\checkmark$ &
  \multicolumn{1}{c|}{} &
  $\checkmark$ \\ \midrule
UA-FedRec~\cite{26yi2023ua} &
  \multicolumn{1}{c|}{$\checkmark$} &
   &
  \multicolumn{1}{c|}{$\checkmark$} &
  \multicolumn{1}{c|}{} &
   &
  \multicolumn{1}{c|}{} &
  $\checkmark$ &
  \multicolumn{1}{c|}{} &
  $\checkmark$ \\ \midrule
SGLD~\cite{01li2016data} &
  \multicolumn{1}{c|}{$\checkmark$} &
  $\checkmark$ &
  \multicolumn{1}{c|}{$\checkmark$} &
  \multicolumn{1}{c|}{} &
  $\checkmark$ &
  \multicolumn{1}{c|}{$\checkmark$} &
   &
  \multicolumn{1}{c|}{$\checkmark$} &
   \\ \midrule
GraphAttack~\cite{02fang2018poisoning} &
  \multicolumn{1}{c|}{} &
  $\checkmark$ &
  \multicolumn{1}{c|}{$\checkmark$} &
  \multicolumn{1}{c|}{} &
   &
  \multicolumn{1}{c|}{$\checkmark$} &
   &
  \multicolumn{1}{c|}{$\checkmark$} &
   \\ \midrule
MSOPDS~\cite{39yeh2023planning} &
  \multicolumn{1}{c|}{} &
  $\checkmark$ &
  \multicolumn{1}{c|}{$\checkmark$} &
  \multicolumn{1}{c|}{} &
   &
  \multicolumn{1}{c|}{$\checkmark$} &
   &
  \multicolumn{1}{c|}{$\checkmark$} &
   \\ \midrule
PoisonRec~\cite{09song2020poisonrec} &
  \multicolumn{1}{c|}{} &
  $\checkmark$ &
  \multicolumn{1}{c|}{} &
  \multicolumn{1}{c|}{$\checkmark$} &
   &
  \multicolumn{1}{c|}{$\checkmark$} &
   &
  \multicolumn{1}{c|}{$\checkmark$} &
   \\ \midrule
PSMU(V)~\cite{89yuan2023manipulating} &
  \multicolumn{1}{c|}{} &
  $\checkmark$ &
  \multicolumn{1}{c|}{$\checkmark$} &
  \multicolumn{1}{c|}{$\checkmark$} &
  $\checkmark$ &
  \multicolumn{1}{c|}{} &
  $\checkmark$ &
  \multicolumn{1}{c|}{$\checkmark$} &
  $\checkmark$ \\ \midrule
IPDGI~\cite{83chen2023Adversarial} &
  \multicolumn{1}{c|}{} &
  $\checkmark$ &
  \multicolumn{1}{c|}{$\checkmark$} &
  \multicolumn{1}{c|}{} &
  $\checkmark$ &
  \multicolumn{1}{c|}{$\checkmark$} &
   &
  \multicolumn{1}{c|}{$\checkmark$} &
   \\ \midrule
TDP-CP~\cite{80zhang2022targeted} &
  \multicolumn{1}{c|}{} &
  $\checkmark$ &
  \multicolumn{1}{c|}{$\checkmark$} &
  \multicolumn{1}{c|}{} &
  $\checkmark$ &
  \multicolumn{1}{c|}{$\checkmark$} &
   &
  \multicolumn{1}{c|}{$\checkmark$} &
   \\ \midrule
ARG~\cite{81chiang2023shilling}&
  \multicolumn{1}{c|}{} &
  $\checkmark$ &
  \multicolumn{1}{c|}{$\checkmark$} &
  \multicolumn{1}{c|}{} &
  $\checkmark$ &
  \multicolumn{1}{c|}{$\checkmark$} &
   &
  \multicolumn{1}{c|}{$\checkmark$} &
   \\ \midrule
KGAttack~\cite{21chen2022knowledge} &
  \multicolumn{1}{c|}{} &
  $\checkmark$ &
  \multicolumn{1}{c|}{} &
  \multicolumn{1}{c|}{$\checkmark$} &
   &
  \multicolumn{1}{c|}{$\checkmark$} &
   &
  \multicolumn{1}{c|}{$\checkmark$} &
   \\ \midrule
KG-Rlattack~\cite{36wu2022poisoning} &
  \multicolumn{1}{c|}{} &
  $\checkmark$ &
  \multicolumn{1}{c|}{} &
  \multicolumn{1}{c|}{$\checkmark$} &
   &
  \multicolumn{1}{c|}{$\checkmark$} &
   &
  \multicolumn{1}{c|}{$\checkmark$} &
   \\ \midrule
FedAttack~\cite{20wu2022fedattack} &
  \multicolumn{1}{c|}{$\checkmark$} &
   &
  \multicolumn{1}{c|}{} &
  \multicolumn{1}{c|}{} &
   &
  \multicolumn{1}{c|}{} &
  $\checkmark$ &
  \multicolumn{1}{c|}{} &
  $\checkmark$ \\ \midrule
Infmix~\cite{37wu2023influence} &
  \multicolumn{1}{c|}{} &
  $\checkmark$ &
  \multicolumn{1}{c|}{$\checkmark$} &
  \multicolumn{1}{c|}{} &
  $\checkmark$ &
  \multicolumn{1}{c|}{$\checkmark$} &
   &
  \multicolumn{1}{c|}{$\checkmark$} &
   \\ \midrule
ClusterAttack~\cite{23yu2023untargeted} &
  \multicolumn{1}{c|}{$\checkmark$} &
   &
  \multicolumn{1}{c|}{} &
  \multicolumn{1}{c|}{} &
   &
  \multicolumn{1}{c|}{} &
  $\checkmark$ &
  \multicolumn{1}{c|}{} &
  $\checkmark$ \\ \midrule
ModelExtractionAttack~\cite{14yue2021black} &
  \multicolumn{1}{c|}{} &
  $\checkmark$ &
  \multicolumn{1}{c|}{} &
  \multicolumn{1}{c|}{} &
   &
  \multicolumn{1}{c|}{$\checkmark$} &
   &
  \multicolumn{1}{c|}{$\checkmark$} &
   \\ \midrule
RAPU~\cite{12zhang2021data} &
  \multicolumn{1}{c|}{} &
  $\checkmark$ &
  \multicolumn{1}{c|}{$\checkmark$} &
  \multicolumn{1}{c|}{} &
   &
  \multicolumn{1}{c|}{$\checkmark$} &
   &
  \multicolumn{1}{c|}{$\checkmark$} &
   \\ \midrule
GTA~\cite{40wang2023revisiting} &
  \multicolumn{1}{c|}{} &
  $\checkmark$ &
  \multicolumn{1}{c|}{$\checkmark$} &
  \multicolumn{1}{c|}{} &
   &
  \multicolumn{1}{c|}{$\checkmark$} &
   &
  \multicolumn{1}{c|}{$\checkmark$} &
   \\ \midrule
AutoAttack~\cite{73guo2023targeted} &
  \multicolumn{1}{c|}{} &
  $\checkmark$ &
  \multicolumn{1}{c|}{$\checkmark$} &
  \multicolumn{1}{c|}{} &
   &
  \multicolumn{1}{c|}{$\checkmark$} &
   &
  \multicolumn{1}{c|}{$\checkmark$} &
   \\ \midrule
CD-Attack~\cite{03chen2019data} &
  \multicolumn{1}{c|}{$\checkmark$} &
  $\checkmark$ &
  \multicolumn{1}{c|}{$\checkmark$} &
  \multicolumn{1}{c|}{} &
   &
  \multicolumn{1}{c|}{$\checkmark$} &
   &
  \multicolumn{1}{c|}{$\checkmark$} &
   \\ \midrule
GSA-GANs~\cite{16wang2022gray} &
  \multicolumn{1}{c|}{} &
  $\checkmark$ &
  \multicolumn{1}{c|}{$\checkmark$} &
  \multicolumn{1}{c|}{} &
  $\checkmark$ &
  \multicolumn{1}{c|}{$\checkmark$} &
   &
  \multicolumn{1}{c|}{$\checkmark$} &
   \\ \midrule
AUSH~\cite{07lin2020attacking} &
  \multicolumn{1}{c|}{} &
  $\checkmark$ &
  \multicolumn{1}{c|}{$\checkmark$} &
  \multicolumn{1}{c|}{} &
  $\checkmark$ &
  \multicolumn{1}{c|}{$\checkmark$} &
   &
  \multicolumn{1}{c|}{$\checkmark$} &
   \\ \midrule
LegUP~\cite{32lin2022shilling} &
  \multicolumn{1}{c|}{} &
  $\checkmark$ &
  \multicolumn{1}{c|}{$\checkmark$} &
  \multicolumn{1}{c|}{} &
  $\checkmark$ &
  \multicolumn{1}{c|}{$\checkmark$} &
   &
  \multicolumn{1}{c|}{$\checkmark$} &
   \\ \midrule
PC-Attack~\cite{74zeng2023practical} &
  \multicolumn{1}{c|}{} &
  $\checkmark$ &
  \multicolumn{1}{c|}{$\checkmark$} &
  \multicolumn{1}{c|}{} &
   &
  \multicolumn{1}{c|}{$\checkmark$} &
   &
  \multicolumn{1}{c|}{$\checkmark$} &
   \\ \midrule
LOKI~\cite{10zhang2020practical} &
  \multicolumn{1}{c|}{} &
  $\checkmark$ &
  \multicolumn{1}{c|}{$\checkmark$} &
  \multicolumn{1}{c|}{} &
   &
  \multicolumn{1}{c|}{$\checkmark$} &
   &
  \multicolumn{1}{c|}{$\checkmark$} &
   \\ \midrule
CopyAttack~\cite{08fan2021attacking} &
  \multicolumn{1}{c|}{} &
  $\checkmark$ &
  \multicolumn{1}{c|}{$\checkmark$} &
  \multicolumn{1}{c|}{} &
   &
  \multicolumn{1}{c|}{$\checkmark$} &
   &
  \multicolumn{1}{c|}{$\checkmark$} &
   \\ \midrule
ReverseAttack~\cite{24zhang2021reverse} &
  \multicolumn{1}{c|}{$\checkmark$} &
  $\checkmark$ &
  \multicolumn{1}{c|}{$\checkmark$} &
  \multicolumn{1}{c|}{} &
   &
  \multicolumn{1}{c|}{$\checkmark$} &
   &
  \multicolumn{1}{c|}{$\checkmark$} &
   \\ \midrule
AIA~\cite{06tang2020revisiting} &
  \multicolumn{1}{c|}{} &
  $\checkmark$ &
  \multicolumn{1}{c|}{$\checkmark$} &
  \multicolumn{1}{c|}{} &
   &
  \multicolumn{1}{c|}{$\checkmark$} &
   &
  \multicolumn{1}{c|}{$\checkmark$} &
   \\ \midrule
SUI-Attack~\cite{72huang2023single} &
  \multicolumn{1}{c|}{} &
  $\checkmark$ &
  \multicolumn{1}{c|}{$\checkmark$} &
  \multicolumn{1}{c|}{$\checkmark$} &
   &
  \multicolumn{1}{c|}{$\checkmark$} &
   &
  \multicolumn{1}{c|}{$\checkmark$} &
   \\ \midrule
AdvAttack~\cite{05christakopoulou2019adversarial} &
  \multicolumn{1}{c|}{} &
  $\checkmark$ &
  \multicolumn{1}{c|}{$\checkmark$} &
  \multicolumn{1}{c|}{} &
  $\checkmark$ &
  \multicolumn{1}{c|}{$\checkmark$} &
   &
  \multicolumn{1}{c|}{$\checkmark$} &
   \\ \midrule
TrialAttack~\cite{13wu2021triple}&
  \multicolumn{1}{c|}{} &
  $\checkmark$ &
  \multicolumn{1}{c|}{$\checkmark$} &
  \multicolumn{1}{c|}{} &
  $\checkmark$ &
  \multicolumn{1}{c|}{$\checkmark$} &
   &
  \multicolumn{1}{c|}{$\checkmark$} &
   \\ \midrule
UBA~\cite{c5wang2024uplift}&
  \multicolumn{1}{c|}{} &
  $\checkmark$ &
  \multicolumn{1}{c|}{$\checkmark$} &
  \multicolumn{1}{c|}{$\checkmark$} &
   &
  \multicolumn{1}{c|}{$\checkmark$} &
   &
  \multicolumn{1}{c|}{$\checkmark$} &
   \\ \midrule
TextRecAttack~\cite{c4zhang2024stealthy}&
  \multicolumn{1}{c|}{} &
  $\checkmark$ &
  \multicolumn{1}{c|}{} &
  \multicolumn{1}{c|}{} &
   $\checkmark$ &
  \multicolumn{1}{c|}{$\checkmark$} &
   &
  \multicolumn{1}{c|}{$\checkmark$} &
   \\ \midrule

RecUP~\cite{17zhang2021attacking} &
  \multicolumn{1}{c|}{} &
  $\checkmark$ &
  \multicolumn{1}{c|}{$\checkmark$} &
  \multicolumn{1}{c|}{} &
  $\checkmark$ &
  \multicolumn{1}{c|}{$\checkmark$} &
   &
  \multicolumn{1}{c|}{$\checkmark$} &
   \\ \bottomrule
\end{tabular}}
\label{summary of poisoning attacks}
\end{table*}

\subsubsection{Hybrid-Goal Attacks}
Attackers might design more flexible attack strategies, aligning with multiple objectives and including both system degradation and targeted manipulation. \textit{SGLD}~\cite{01li2016data} stands out as a pioneering method capable of pursuing dual goals concurrently. Its core technique involves maximizing the discrepancy in all predictions before and after the poisoning attack (system degradation) and enhancing the prediction likelihood of a target item for each user (targeted manipulation). It achieves a balance between these two goals using weighted coefficients. Following the groundwork laid by \textit{SGLD}, studies such as \textit{CD-Attack}~\cite{03chen2019data} and \textit{NCFAttack}~\cite{25zhang2020towards} have adopted this dual-goal framework. \textit{CD-Attack} demonstrates that malicious profiles developed using this methodology exhibit good transferability across different domains, while \textit{NCFAttack} chooses to customize a more elegant attack target. It is different from the former, which takes the attack effect as the optimization goal. Instead, it generates specific adversarial gradients based on the model optimization strategy to enable the execution of more precise attacks.

\stitle{Remark.} System degradation and targeted manipulation are two distinct and clearly defined attack objectives in RS. However, the potential for hybrid attacks, which combine these goals, has not been fully explored. From an attacker's standpoint, orchestrating an attack that simultaneously achieves both objectives usually involves careful tuning of hyperparameters to find a balance, treating the goals as if they hinder each other. This approach can be cumbersome and overlooks the possibility that these objectives might actually complement and enhance one another. Moreover, current attack methodologies driven by attack goals tend to converge towards similar objectives, which might not fully represent the diversity of real-world attackers' intentions. This gap in the strategic landscape suggests a significant opportunity for attackers to develop and employ more varied and complex attack strategies, indicating a pressing need for further exploration in this area.

\subsection{Attacks Limited by Practical Capability}
To execute a successful poisoning attack on real-world RS, attackers should consider the \textbf{knowledge constraint}, encompassing the training model and data utilized in RS, which affects vulnerability exploration. Additionally, they must operate within \textbf{cost limitations}, including budgets for injecting malicious users or querying the RS. Ensuring the \textbf{invisibility} of the poisoned data is also essential to avoid detection.

\subsubsection{Knowledge Constraint}
In RS model-unknown contexts, attackers often train and employ a surrogate model to deduce representations of RS model, adapting their strategies to the constraints of their knowledge about the victim system. For example, ModelExtractionAttack~\cite{14yue2021black} employs model distillation techniques, designating the victim black box model as the teacher and the surrogate model as the student. The surrogate model is trained to produce outputs that are as identical as possible to those of the victim black box model. Following the model distillation concept, \textit{PC-Attack}~\cite{74zeng2023practical} further establishes two objectives: ensuring the representation of the target item is as similar as possible to that of other items, and making the representation of users who interacted with the target item as similar as possible to other users. It then employs a ``pre-training and fine-tuning" strategy, where a simulator model is pre-trained on multiple data sources, such as public recommendation datasets. A small portion of the target data is then input into the simulator model for fine-tuning.

In addition to surrogate models, reinforcement learning (RL)~\cite{c11arulkumaran2017deep} is also extensively utilized in model-unknown contexts. Attackers design appropriate reward functions for RL to optimize the attack effect without directly using the victim model to provide simulated representations of RS. For example, \textit{LOKI}~\cite{10zhang2020practical} implements RL through designing the reward as the weighted averaged influence on the prediction scoring function, facilitating the generation of user behavior samples, while \textit{CopyAttack}~\cite{08fan2021attacking} employs RL using a ranking scoring function, which grants a positive reward if the target item appears in the recommended list. This method identifies and replicates real user profiles from a source domain into the target domain.

The aforementioned poisoning methods are generally agnostic to data-unknown contexts. However, it is also necessary to overcome the limitations imposed by varying levels of knowledge about the victim data. Some poisoning attacks operate under the assumption of minimal knowledge of the training set. For instance, \textit{AIA}\cite{06tang2020revisiting} uses limited user-item interactions as the basis information and leverages a well-trained surrogate model to infer additional interactions. The inferred data is then used to refine the attack optimization process, increasing the attack's destructiveness. Continuing the strategy of using surrogate models for data inference, \textit{ReverseAttack}\cite{24zhang2021reverse} collects social information datasets to aid in the inference. On the other hand, some methods cater to entirely data-free scenarios where no direct access to training data is available. Under such conditions, \textit{A-hum}~\cite{22rong2022poisoning} hypothesizes that the directions of users' embedding vectors should be uniformly distributed. It models a benign user's embedding vector as a vector-valued random variable, approximating the likely characteristics of these vectors based on statistical assumptions. Delving deeper into data-constrained environments, \textit{RAPU}~\cite{12zhang2021data} tackles the additional challenges presented by data obscurities and noise. It employs a bi-level optimization framework paired with a probabilistic generative model to manage data incompleteness and perturbations effectively. This sophisticated approach allows for a more accurate estimation of user and item characteristics, substantially improving the attack's effectiveness by compensating for the lack of direct data access.

\subsubsection{Cost Limitation}
Given that each instance of manipulated or injected malicious data incurs a certain cost, which is naturally limited by the attackers' financial resources, optimizing poisoning attacks to maximize their effectiveness while minimizing the associated costs is a crucial goal for attackers. For example, \textit{SUI-Attack}~\cite{72huang2023single} addresses the constraint of a limited budget for injecting malicious users. It explores a scenario in which an attacker can only inject a single user. Despite this limitation, this approach remains effective by carefully selecting suitable items for interaction by the injected user, thereby showcasing the potential of executing poisoning attacks under tight constraints.

Some poisoning attacks consider scenarios with a limited budget for system output queries. \textit{ModelExtractionAttack}~\cite{14yue2021black} employs knowledge distillation to convert opaque, black-box victim models into transparent, white-box surrogate models that are more easily manipulable. This transformation allows the replacement of the victim models to fulfill query tasks within budgetary constraints. Following a similar theme of limited query budgets, \textit{CopyAttack}~\cite{08fan2021attacking,71fan2023adversarial} pre-trains a surrogate model using interaction data from querying other multiple platforms, which aims to identify properties that are both inherent and transferable across various platforms. Subsequently, the model is fine-tuned with partial target data, customizing it for specific attack objectives.

\subsubsection{Invisibility}
To protect RS from poisoning attacks, most RS implementations are equipped with detection mechanisms for identifying such attacks. As a result, effectively executing poisoning attack methods requires meticulous attention to the invisibility of malicious profiles, ensuring they remain undetected by these protective systems.

Adversarial learning is a key technology in ensuring the invisibility of malicious user profiles, achieved by aligning the distribution of adversarial fake users with that of real user interactions. \textit{AdvAttack}~\cite{05christakopoulou2019adversarial} is the pioneering study that integrates adversarial learning into poisoning attacks. Building upon such an idea, subsequent studies have utilized the more advanced framework of Generative Adversarial Networks (GANs)~\cite{58goodfellow2020generative}, comprising generative and discriminative components. These components engage in a competitive interaction to enhance the resemblance of generated user profiles to real user profiles. Studies such as \textit{AUSH}~\cite{07lin2020attacking} and LegUP~\cite{32lin2022shilling} focus on refining the generative component to improve its efficacy. Conversely, another line of research has concentrated on enhancing the discriminative components. For instance, \textit{TrialAttack}~\cite{13wu2021triple} incorporates an additional discriminator to assess the impact of the synthesized user profiles, offering a more detailed evaluation of the attack's influence within RS.

Beyond adversarial learning, other methods also consider profile invisibility. For example, \textit{PipAttack}~\cite{11zhang2022pipattack} requires minimal impact on the overall recommendation performance during targeted manipulation attacks. \textit{FedRecAttack}~\cite{19rong2022fedrecattack} imposes limits on the maximum gradient perturbations of target items post-attack, as significant perturbations could alert defenders. In a more direct approach, \textit{RecUP}~\cite{17zhang2021attacking} and \textit{GSA-GANs}~\cite{16wang2022gray} employ state-of-the-art malicious user detection methods to evaluate whether generated fake users can evade detection mechanisms.

\stitle{Remark.}
As attackers, understanding and leveraging one's prior knowledge and attack budget, while ensuring the invisibility of the attack, can significantly enhance the effectiveness of poisoning attacks. The current methods for the practical application of poisoning attacks have reached a relatively mature stage. However, the assumptions made about the attackers' capabilities in research are often set at a higher level than what is realistic. In reality, attackers usually have weaker capabilities. For example, the attack budget is often set at 1\% of the original dataset for controlled users, which represents a considerable expenditure. Moreover, attackers face stringent detection mechanisms when scraping training data, limiting the amount of data they can realistically obtain. Therefore, further exploration of attackers' capabilities under more stringent constraints is essential to discover practical methods that real-world attackers might employ when executing attacks.

\subsection{Attacks Specific to Victim Architecture}
For an attacker, comprehending the distinctions between \textbf{centralized recommendation} and \textbf{decentralized recommendation} systems is crucial. Centralized systems may be compromised through a single, high-impact intervention, whereas decentralized systems often require a more covert, prolonged effort.

\subsubsection{Centralized Recommendation}
Starting with traditional centralized systems, studies such as \textit{TNA}\cite{34fang2020influence} explore the vulnerabilities of matrix factorization-based RS, while \textit{UNAttack}\cite{31chen2021data} investigates neighborhood-based systems. Both methods construct hypothetical rating matrices post-malicious file injection and iteratively optimize these matrices to evaluate and identify more threatening attack vectors. As research progresses, it extends to more complex deep learning-based RS. \textit{DLAttack}\cite{18huang2021data} targets pure deep learning-based RS, and \textit{NCFAttack}\cite{25zhang2020towards} focuses on neural collaborative filtering-based RS. Both approaches develop rigorous models to specify attack constraints within their recommendation architecture and define various objective functions to capture the fundamental goals of availability and targeted attacks. Furthermore, \textit{GOAT}~\cite{15wu2021ready} and \textit{GSPAttack}~\cite{33nguyen2022poisoning} execute poisoning attacks specifically on graph-based RS that predominantly incorporate GNN~\cite{60zhang2019graph} in their architectures. These studies expose vulnerabilities in the gradient update mechanisms of these DL-based systems. 
This evolution continues with the advent of Self-Supervised Learning (SSL) ~\cite{41yu2023self}. \textit{SSLAttack}~\cite{30wang2023poisoning} targets this paradigm, shifting the focus of the poisoning attacks to the initial pre-training phase. This approach stresses the injection of adversarial influences during pre-training, which are subsequently propagated to downstream recommendation models during the fine-tuning stage. In a similar context, \textit{PromptAttack}~\cite{27wu2023attacking} explores vulnerabilities in pre-trained models utilizing prompt learning~\cite{c13zhou2022conditional}, a nuanced approach compared to traditional fine-tuning, thus exposing weaknesses in various downstream recommendation models. 
\textit{CLeaR}~\cite{82wang2023poisoning} also emphasizes the focus on SSL, intensifying contrastive learning loss in RS can result in a uniform distribution of representations. 
Additionally, in response to the growing demand for transparency, explainable RS, which converts user interactions into logical expressions, has been scrutinized.
\textit{TextRecAttack}~\cite{c4zhang2024stealthy} highlights the security risks associated with integrating large language models (LLMs)~\cite{c14fan2023recommender} into RS. This attack illustrates how minor textual modifications during the test phase can significantly enhance an item's exposure, circumventing the need for direct manipulation of the model’s training infrastructure.
\textit{H-CARS}~\cite{28chen2023dark} leverages counterfactual explanations to craft malicious user profiles, exploiting the logical reasoning in these systems and thus uncovering new vulnerabilities. 

\subsubsection{Decentralized Recommendation}
As concerns over privacy in centralized systems grew, research pivoted towards decentralized federated RS, where local storage of user data enhances security and privacy. 
Despite the enhanced robustness of this architecture, \textit{FedRecAttack}~\cite{19rong2022fedrecattack} illustrates that public interactions in federated RS can be exploited to approximate user features. These derived user features can subsequently be utilized to intensify the manipulation of recommendations.
\textit{PSMU}~\cite{29yuan2023manipulating} identifies a high similarity between the top-K recommendations for randomly generated synthetic users and actual users. This observation indicates that attackers can manipulate synthetic users to obtain feedback on targeted items, and this feedback can subsequently be leveraged to refine the attack model.
Building on the above work, \textit{A-hum}~\cite{22rong2022poisoning} further focuses on identifying hard users who consider the target items as negative samples and manipulate malicious interactions to elevate these users' preferences for the targeted items.
\textit{PipAttack}~\cite{11zhang2022pipattack} capitalizes on the distinctive characteristics of popular item aggregation within the widely used BPR loss. It alters the target item to resemble the attributes of popular items in the embedding space, thereby increasing the likelihood of user interactions with that item. \textit{UA-FedRec}~\cite{26yi2023ua} modifies the representations to increase the distance between similar samples and decrease it for dissimilar ones to standardize the distances across all user and item representations. This deliberately created uniform distribution can disrupt each user's recommendation results and degrade overall recommendation performance.

\stitle{Remark.}
Different victim architectures exhibit unique vulnerabilities, and once attackers identify these weaknesses, they can conduct targeted poisoning attacks with significantly enhanced effectiveness. However, the continuous evolution and iteration of victim architectures can constrain the application of specific attack methods. Since the development of attack methods inevitably lags behind the mainstream advancements in the recommendation community, attackers may be more inclined to explore unified attack methods to adaptively ensure the effectiveness of their attacks. This adaptability is crucial in maintaining the potency of attacks across varying architectural paradigms.

\subsection{Attacks Divided by Poisoning Strategies}
Attackers can infiltrate systems through two main ways: \textbf{data poisoning} and \textbf{model poisoning}. Data poisoning involves inserting malicious interaction or multimodal data into the RS, thereby contaminating the training dataset with imperceptible alterations. On the other hand, model poisoning disrupts the learning process by altering model gradients. This method entails generating and strategically uploading compromised gradients during the aggregation phase, where data is gathered from multiple user devices.
\subsubsection{Data Poisoning}
In the study of data poisoning attacks, the basic input is user-item interaction data. \textit{SGLD}~\cite{01li2016data} stands out as a pioneering method to explore the vulnerability of interaction data, which merely creates malicious user-item interaction in the matrix factorization-based system to achieve the target item promotion task successfully. Advancing from the general form of interaction data, graph-based RS introduces a significant shift, where users and items become nodes, connected by interactions as edges. The graph-specific poisoning attack methods like \textit{GraphAttack}~\cite{02fang2018poisoning} and \textit{GSPAttack}~\cite{33nguyen2022poisoning} demonstrate vulnerabilities inherent to graph structures. They exploit item recommendation rates as their objective function, optimizing malicious user-item interactions by leveraging the feature of graph propagation dynamics. The complexity further escalates with the incorporation of heterogeneous data, such as social relationships. \textit{MSOPDS}~\cite{39yeh2023planning}, for example, navigates the simultaneous strategies that optimize user-item and user-user interactions of multiple attackers, each with unique social connections. Another type of interaction data is temporal data, pivotal in sequential RS, and also presents a viable target for a poisoning attack. \textit{PoisonRec}~\cite{09song2020poisonrec} employs RL to analyze user preferences from historical sequences, conceptualizing this process as a Markov Decision Process (MDP)~\cite{c15garcia2013markov}. This RL application facilitates the execution of poisoning attack tasks by leveraging the sequential nature of user interactions.

In scenarios involving multimodal data, the specific vulnerabilities to poisoning attacks become more evident. \textit{PSMU}(V)~\cite{89yuan2023manipulating} and \textit{IPDGI}~\cite{83chen2023Adversarial} use guided diffusion models to create adversarial images aimed at facilitating item promotion. These models capably emulate the distribution of benign images, resulting in adversarial images that closely resemble the original ones in terms of fidelity. Furthermore, methods like \textit{TDP-CP}~\cite{80zhang2022targeted} and \textit{ARG}~\cite{81chiang2023shilling} concentrate on the manipulation of textual data associated with items. They implement methods like adding, deleting, replacing words, or even generating new textual content with the objective of inducing shifts in recommendations. More modal information is incorporated with the introduction of KGs~\cite{86ji2021survey}. These elements add detailed entity attributes, such as user demographics and item descriptions. \textit{KGAttack}~\cite{21chen2022knowledge} and \textit{KG-RLattack}~\cite{36wu2022poisoning} utilize the extensive information in KGs to create synthetic user profiles that are both authentic and credible. These poisoning attack methods illustrate the extensive range of poisoning attacks across different data modalities.

\subsubsection{Model Poisoning}
Up to now, model poisoning attacks have mainly appeared in decentralized RS. For example, \textit{PipAttack}~\cite{11zhang2022pipattack} involves a few malicious users uploading meticulously crafted gradients during the model's update process. This allows the target item to adopt the attributes of popular items within the embedding space. \textit{FedRecAttack}~\cite{19rong2022fedrecattack} makes use of the public interactions to approximate users' feature vectors, and then design a bi-level optimization technique to generate poisoned gradients accordingly and control malicious users to upload the poisoned gradients in a well-designed way. \textit{A-hum}~\cite{22rong2022poisoning} optimizes the presentation by gradient descent to mine hard users for a target item, and then uploads poisoned gradient into the server. 
\textit{FedAttack}~\cite{20wu2022fedattack} adopts a strategy in which it selects items that are most and least relevant to user embeddings as the hardest negative and positive samples, respectively. Malicious clients then generate and upload poisoned gradients designed to swap the model's predictions for these two samples, thereby confusing the model's output.

\stitle{Remark.}
Data poisoning attacks and model poisoning attacks exhibit clear differences in their approaches and their effects on the system, making it relatively straightforward to choose between them based on the attack objective. However, it is important to note that model poisoning attacks require the uploading of poisoned gradients, making them primarily infeasible in centralized recommender systems. On the other hand, data poisoning attacks can occur in both decentralized and centralized scenarios. For instance, PSMU(V) ~\cite{89yuan2023manipulating} involves constructing and uploading adversarial images poisoned by synthetic malicious users in a decentralized recommendation setting.

\section{Defense against Poisoning Attacks}
Defense mechanisms against poisoning attacks in RS can be divided into two primary categories: \textbf{poisoning data filtering} and \textbf{robust training}. Poisoning data filtering defenses are implemented prior to the infiltration of malicious elements into the victim RS. This category can be systematically subdivided into three primary methods: supervised anomaly filtering, unsupervised clustering filtering, and confidence-based filtering. 
Robust training defenses are implemented during the training phase if malicious elements have entered the victim RS and can be divided into three types: collaborative enhancement training, confidence-aware training, and contextual enrichment training. The first two types focus on enhancements at the model level, while the latter one emphasizes improvements at the data level. The main technologies used, along with their respective categories, are summarized in Table \ref{combined defense methods}.

\subsection{Poisoning Data Filtering}
Defenders employ various techniques to filter out poisoned data before it is utilized within the system. \textbf{Supervised anomaly filtering} establishes a boundary in the vector space to differentiate between normal and malicious data, enabling proactive filtering of harmful elements. \textbf{Unsupervised clustering filtering} detects and removes anomalies by identifying clusters that significantly deviate in size or composition from the norm. \textbf{Confidence-based filtering} uses probabilistic or statistical models to evaluate and score each data point based on its likelihood of being malicious, filtering out entries with high anomaly scores to maintain the system's integrity.

\subsubsection{Supervised Anomaly Filtering}
Supervised anomaly filtering methods differentiate various categories of samples by establishing clear separating hyperplanes. Traditional detection methods often rely on statistical learning techniques such as Logistic Regression (LR)~\cite{c16lavalley2008logistic}, Support Vector Machines (SVM)~\cite{c17zhang2014hht}, and tree-based models~\cite{c18yang2016re}.
One prevalent approach within such supervised methods involves augmenting traditional detection techniques. For example, the \textit{TSA-TF} model~\cite{d18xu2019detecting} introduces a rating distribution prediction framework tailored specifically to detect suspicious items. This model incorporates four unique features extracted from rating patterns and trust relationships. Subsequently, an SVM classifier is trained to differentiate attack profiles from a set of suspicious user profiles, effectively improving the detection capabilities beyond traditional methods.

While traditional detection methods relying on statistical learning were prevalent, contemporary trends in detection technology have shifted towards harnessing advanced encoding techniques to extract latent features distinguishing genuine from malicious profiles. This shift reflects a broader movement in the field, emphasizing the use of sophisticated neural architectures like Gated Recurrent Units (GRU)~\cite{c19chung2014empirical}, attention mechanisms~\cite{c20vaswani2017attention}, and AutoEncoder~\cite{c25tschannen2018recent}. For example, \textit{RLDetection}~\cite{d5cao2020adversarial} combines attention mechanisms and GRU to create a detection model that distinguishes between normal and adversarial examples in RL-based interactive RS. Similarly, \textit{SDRS}~\cite{d14xu2019slanderous} employs a Hierarchical Dual-Attention recurrent Neural network with a modified GRU to uncover hidden opinions in reviews. It introduces a joint filtering method that detects discrepancies between ratings and reviews to identify malicious profiles. \textit{RecMR}~\cite{d2hsiao2022unsupervised} explores using an AutoEncoder as a detection encoding model, which enhances embeddings through joint training tasks for user recommendations and then applies these enhanced embeddings for binary detection of anomalies.

To delve deeper into feature extraction, researchers are increasingly leveraging graph structures over traditional data formats. \textit{NFGCN-TIA}\cite{d13wang2022detecting} conceptualizes the dataset as a graph, with users as nodes and co-rating relations as edges. By assigning weights to these edges and filtering out typical user relations, it constructs a graph that predominantly highlights suspicious user interactions and employs a three-layer GCN\cite{c22zhang2019graph} model to encode such a graph for detecting malicious users. Building on this graph-based framework, \textit{DHAGCN}~\cite{d15hao2023detection} advances the methodology by further utilizing GCNs. Specifically, it extracts five distinct user features from sequences of item popularity and rating values, clustering user nodes into several groups. These clusters then serve as enhanced training samples for a GCN-based detector to complete the detection task.

\subsubsection{Unsupervised Clustering Filtering}
Unsupervised clustering filtering methods are particularly valuable in scenarios where labels are scarce, employing techniques that can discern patterns and group similar data points without prior categorization. One mainstream is to combine malicious context hidden in recommendations into improved K-Means algorithms. 
For example, \textit{ARBD}~\cite{d9yu2019robust} utilizes a modified K-Means algorithm that leverages two specific metrics: the popularity degree of an item and the distance between users. By adjusting the clustering criteria to these particular dimensions, ARBD enhances its ability to detect outlier activities more effectively.
\textit{GAGE}~\cite{d19zhang2020graph} extends the utility of clustering by constructing a user relationship graph based on user rating behaviors. It employs graph embedding techniques to transform the complex relationships into low-dimensional vector representations of each node (user). Following this transformation, the K-Means++ clustering algorithm is applied to these vectors to form groups and facilitate the identification of malicious activities.
\textit{GSD}~\cite{d10zhang2020detecting} introduces a temporal element into the clustering process by incorporating time sequential information. GSD analyzes the scoring trajectory of each item, segmenting these trajectories based on fixed time intervals to form candidate groups. Then the bisection k-Means algorithm is used to further classify these groups based on their degree of suspicion, effectively isolating the groups likely involved in poisoning attacks.

Aside from the well-known K-Means algorithm, several other clustering techniques have demonstrated their effectiveness in detecting anomalies within various systems. These methods leverage diverse approaches to clustering, utilizing spectral clustering, density-based clustering, and even hierarchical clustering to enhance detection capabilities.
\textit{BS-SC}~\cite{d11cai2019bs} explores the capabilities of spectral clustering. This method manually defines multiple types of features to construct a user behavior similarity matrix and then reconstructs the behavior graph based on rank computation. By utilizing the spectral clustering algorithm, it effectively clusters malicious profiles based on their behavioral similarities. \textit{DSA-AURB}~\cite{d17cai2019detecting} employs a density-based clustering approach. Initially, it identifies target items and the intentions of attacking users by analyzing deviations in rating patterns for each item. The analysis extends to users' rating behaviors from an interest preference perspective, using entropy and block entropy metrics to measure the diversity and memory of users' preferences. These combined metrics are then used as the density information for clustering. 

Another approach involves enhancing basic clustering algorithms with additional components to improve clustering performance. For instance, \textit{MMD}~\cite{d12xu2020detect} integrates metric learning and clustering. It begins by converting reviews into sentiment scores, followed by a sophisticated malicious user profiling component that identifies discrepancies between sentiment scores and ratings. This method incorporates these discrepancies to derive a suitable metric matrix, which is then added to the clustering approach for detection. Similarly, \textit{UD-HMM}~\cite{d20zhang2018ud} combines a hidden Markov model (HMM) with hierarchical clustering to detect poisoning attacks. It models users' historical rating behaviors using HMM and evaluates each user's level of suspicion by analyzing their preference sequences and rating behavior differences compared to genuine users. A hierarchical clustering method then groups users based on their suspicious degrees.

\subsubsection{Confidence-Based Filtering}
Confidence-based filtering methods investigate probabilistic inference and trustworthiness evaluation of malicious profiles and then decide if the profiles are malicious based on confidence. 
\textit{PRD-BGP}~\cite{d3liu2020recommending} investigates various combination strategies for interaction data to construct a comprehensive interaction matrix. This method evaluates pairwise ranking relations across all entities within the unified matrix and formulates a novel pairwise loss objective function to assess the malicious degree of instances, where higher rankings may indicate malicious results. Meanwhile, \textit{RICD}~\cite{d1li2021large} utilizes a risk probability detection approach based on the premise that users who interact with ordinary items yet frequently click on popular or `hot' items might be exhibiting suspicious behavior. This method initially identifies a group of potentially suspicious users, and then refines this group through further scrutiny to assess the risk associated with user and item behaviors, ultimately determining malicious intent based on high-risk scores. 

The aforementioned methods leverage the interaction patterns to identify suspicious behaviors, which, while insightful, often struggle with the issue of data sparsity. To address this limitation, some detection methods propose leveraging attribute data to enhance the detection of malicious activities.
For instance, \textit{PITA} \cite{d6yang2020probabilistic} uses probabilistic inference derived from behavioral links, which are examined within a coupled association network built from user rating behaviors and item attributes. It employs a factor graph model to evaluate the reliability of link behaviors, identifying and highlighting potentially malicious links.
Analogously, \textit{SAD}~\cite{d7aktukmak2021sequential} focuses on the detection within the context of sequential RS. It integrates rating history with mixed attribute data and employs a latent variable model trained via a variational EM (Expectation-Maximization) algorithm, facilitating the detection of anomalies in user sequences.
\textit{QADetection}~\cite{d4aktukmak2019quick}, on the other hand, proposes a probabilistic generative model that integrates user attributes and observed ratings into a latent space to generate anomaly statistics for identifying potentially malicious users.

Building on these approaches, some methods further explore unique scenarios to address specific challenges or new conditions. For instance, \textit{PDR}~\cite{d8lai2023towards} aims to address the problem of noisy data labels, encompassing both benign noise and malicious poisoned data. This method employs a strategy that dynamically adjusts prior knowledge to refine and enhance the detection of malicious profiles. This adaptive approach allows PDR to continuously update and improve its detection capabilities as more data becomes available or as user behaviors evolve. Additionally, \textit{FSAD}~\cite{d22jiang2020detection} pioneers the study of malicious attacks within the context of federated learning. FSAD innovatively designs four novel features derived from the gradients exchanged among clients during the learning process. By leveraging these gradient-based features, FSAD trains a semi-supervised Bayes classifier using an EM algorithm to effectively identify malicious attackers.

\stitle{Remark.}
Poisoning data filtering is a critical process for defenders to complete their defense mechanisms, involving the meticulous examination of data before its incorporation into model training or RS. However, there remains a significant gap in the field of poisoning data filtering: the current techniques are fragmented and fail to seamlessly integrate with advanced technological paradigms, thereby greatly limiting their efficacy in protecting RS. With the rising prominence of data-centric AI \cite{zha2023data}, there is a compelling opportunity to revolutionize poisoning data filtering. Data-centric AI focuses on improving data quality and system robustness, making it ideally suited to enhance anti-poisoning measures. By developing filtering techniques that are not only robust but also deeply integrated with data-centric AI principles, we can substantially elevate our ability to detect and neutralize sophisticated poisoning attacks. This advancement is crucial for maintaining the integrity and reliability of RS, ensuring they are secure against evolving threats and continue to function as trusted components of our technological infrastructure.

\subsection{Robust Training}
Robust training employs strategies to enhance the resilience of RS. \textbf{Collaborative enhancement training} exploits the synergy of multiple models, using adversarial or cooperative methods to improve each model's resistance to attacks. \textbf{Confidence-aware training} dynamically evaluates the trustworthiness of each sample, emphasizing learning from reliable data and reducing the impact of potentially poisoned inputs. \textbf{Contextual enrichment training} enriches the training dataset with a diverse range of additional signals. This enrichment strengthens the model’s contextual understanding, enabling it to detect and withstand attacks more effectively by making more informed and nuanced predictions.
\begin{table*}[]
\caption{A summary of poisoning data filtering and robust training defense methods.} 
\centering
\resizebox{1\textwidth}{!}{
\begin{tabular}{@{}c|c|c@{}}
\toprule
\textbf{Model}   & \textbf{Main Technologies}     & \textbf{Strategy} \\ \midrule
TSA-TF~\cite{d18xu2019detecting}            & SVM                         & Supervised Anomaly Filtering      \\
RLDetection~\cite{d5cao2020adversarial}, SDRS~\cite{d14xu2019slanderous},RecMR~\cite{d2hsiao2022unsupervised},NFGCN-TIA~\cite{d13wang2022detecting}, DHAGCN~\cite{d15hao2023detection}  & Attention, GRU, AutoEncoder, GNN                & Supervised Anomaly Filtering      \\
DSSD-ImMPL~\cite{d22jiang2020detection}       & Model Distillation          & Supervised Anomaly Filtering      \\
ARBD~\cite{d9yu2019robust}, GAGE~\cite{d19zhang2020graph}, GSD~\cite{d10zhang2020detecting}    & K-Means                     & Unsupervised Clustering Filtering        \\
MMD~\cite{d12xu2020detect}              & Metric Learning             & Unsupervised Clustering Filtering        \\
BS-SC~\cite{d11cai2019bs}            & Spectral Clustering         & Unsupervised Clustering Filtering        \\
DSA-AURB~\cite{d17cai2019detecting}         & Density-Based Clustering    & Unsupervised Clustering Filtering        \\
UD-HMM~\cite{d20zhang2018ud}           & Hierarchical Clustering     & Unsupervised Clustering Filtering        \\
PRD-BGP~\cite{d3liu2020recommending}          & Pairwise Learning           & Confidence-Based Filtering     \\
RICD~\cite{d1li2021large}, PITA~\cite{d6yang2020probabilistic}, PDR~\cite{d8lai2023towards},SAD~\cite{d7aktukmak2021sequential}, FSAD~\cite{d22jiang2020detection}     & Risk Probability Prediction & Confidence-Based Filtering     \\
QADetection~\cite{d4aktukmak2019quick}       & Probabilistic Generative    & Confidence-Based Filtering     \\
APT~\cite{a26wu2021fight}, APR~\cite{a35he2018adversarial}, PDAug~\cite{a13wang2023robust}, AHGNNRec~\cite{a32sang2023adversarial}, AMR~\cite{a34tang2019adversarial} & Adversarial Learning         & Collaborative Enhancement Training              \\
DR~\cite{a6wang2019doubly}, StableDR~\cite{a5li2022tdr}, MR~\cite{a5li2022tdr}, CDR~\cite{a7sun2024theoretically}, BirDRec~\cite{a7sun2024theoretically} & Doubly Robust                & Collaborative Enhancement Training              \\
ADT~\cite{a31wang2021denoising}, HiCS~\cite{29yuan2023manipulating}, KRDN~\cite{28chen2023dark}              & Truncated Loss               & Confidence-Aware Training      \\
ADT~\cite{a31wang2021denoising}, SGDL~\cite{a27gao2022self}, BOD~\cite{42wang2023efficient}, DeCA~\cite{a9zhang2023denoising}, AutoDenoise~\cite{a8lin2023autodenoise} & Reweighted Loss                    & Confidence-Aware Training      \\
DUMN~\cite{a22bian2021denoising}, GDMSR~\cite{a12quan2023robust}, DPT~\cite{a9zhang2023denoising}                & Connection Information Enhancement & Contextual Enrichment Training \\
Conde~\cite{a15liu2021concept}, AiD~\cite{29yuan2023manipulating}, LoRec~\cite{c3zhang2024lorec}                     & Modal Information Enhancement      & Contextual Enrichment Training \\
DPGN~\cite{a16xu2023dpgn}                       & Time Information Enhancement & Contextual Enrichment Training \\ \bottomrule
\end{tabular}}
\label{combined defense methods}
\end{table*}

\subsubsection{Collaborative Enhancement Training}
Models can enhance their robustness against attacks by incorporating more interactive modalities and employing collaborative learning to enhance informative signals. A potent approach within this paradigm is the adversarial architecture, which typically involves one component of the model simulating the role of an attacker, introducing disturbances in data or representations to strengthen the system against actual malicious activities. The other main stream is double robustness architecture~\cite{a6wang2019doubly}, which jointly trains a rating prediction model and an error imputation model. The rating prediction model focuses on recovering prediction errors for missing ratings, while the error imputation model is designed to weight observed ratings according to their likelihood of being observed. By integrating these components in a doubly robust manner, double robustness achieves unbiased performance estimation and effectively mitigates the impact of propensity variance, enhancing the system’s overall resilience.

As a notable example of adversarial architecture, \textit{APT}~\cite{a26wu2021fight}, adopts a `fighting fire with fire' defense mechanism. It simulates the poisoning process by injecting fake user data to minimize empirical risk, thereby fostering a more robust system capable of resisting genuine attacks.
In contrast, \textit{APR}~\cite{a35he2018adversarial} enhances the robustness by introducing adversarial perturbations at the representation level. This approach resembles playing a minimax game where the goal is to minimize the BPR objective function while simultaneously defending against an adversary who attempts to maximize this objective by injecting adversarial perturbations into the model parameters.
Building on the strengths of both APR and APT, methods like \textit{PDAug}~\cite{a13wang2023robust} and \textit{AHGNNRec}~\cite{a32sang2023adversarial} utilize adversarial perturbations to optimize the RS further. \textit{PDAug} actively modifies original samples in a direction that maximizes the loss function, thereby generating new data from these altered samples for more effective system exploration and training. \textit{AHGNNRec} applies adversarial training to a hierarchical heterogeneous GNN, which enhances its ability to learn user and item embeddings by recognizing the unique contributions of various interaction types.
Additionally, methods like \textit{AMR}~\cite{a34tang2019adversarial} focus on generating adversarial information on modal data, training the system to defend against adversaries who add perturbations to the target image to decrease the model’s accuracy.

Building upon double robustness framework, \textit{StableDR}~\cite{a5li2022tdr} introduces a cycle learning concept which cyclically updates the imputation, propensity, and prediction models, leading to more stable and accurate predictions.
Similarly, \textit{MR}~\cite{a5li2022tdr} extends the concept of model robustness by utilizing multiple candidate imputation and propensity models. Moreover, some methods address special data circumstances that could compromise model integrity. For instance, \textit{CDR}~\cite{a7sun2024theoretically} improves upon the \textit{DR} approach by introducing a conservative Doubly Robust strategy. This strategy scrutinizes both the mean and variance of imputations to shield against targeted poisoning attacks, fortifying the model's defenses.
\textit{BirDRec}~\cite{a7sun2024theoretically} targets scenarios with unreliable inputs or labels. It implements a rectification sampling strategy and a self-ensemble mechanism to effectively purge noisy data during the mutual training process of DR models. This approach not only cleanses the data input but also reinforces the training process, ensuring that the recommendations generated are both accurate and reliable.

\subsubsection{Confidence-Aware Training}
Adaptively adjusting the confidence levels of samples during model training has proven to be a highly effective method to mitigate the effects of poisoning attacks on RS. This strategy enhances model robustness by dynamically modifying how much certain data points influence the training process based on their assessed reliability. \textit{ADT}~\cite{a31wang2021denoising} is a pioneering method in this area, capitalizing on the observation that noisy feedback often generates large loss values during the initial stages of training in typical RS. Drawing inspiration from this insight, ADT introduces a novel training strategy that adaptively prunes noisy interactions. The approach incorporates two adaptive loss formulation paradigms: Truncated Loss: This technique discards samples with large loss values by setting a dynamic threshold in each training iteration. This effectively removes data points likely to be noise or malicious inputs, preventing them from adversely affecting the model.
Reweighted Loss: In contrast, this method does not discard high-loss samples outright but instead reduces their weight in the training process. This allows the model to still learn from these samples but limits their influence, thereby safeguarding against potential poisoning.

Building on the concept of truncated loss, \textit{HiCS}~\cite{29yuan2023manipulating} employs a gradient clipping strategy with adaptive limits. This method sets thresholds based on the average normalization of the largest gradients observed, clipping all gradients that exceed these limits to prevent disproportionate impacts from anomalous data points on the model's learning trajectory.
Further enhancing the approach, \textit{KRDN}~\cite{28chen2023dark} integrates a self-adapted truncated loss function with the rich contextual data provided by knowledge graphs. This strategy not only prunes noisy implicit feedback but also utilizes an adaptive knowledge refining strategy to distill high-quality knowledge graph triplets for aggregation. By doing so, KRDN not only prunes out malicious data but also enriches the training process with high-value, reliable information, significantly enhancing the recommendation quality.

The concept of reweighted loss, aimed at adaptively managing the influence of training data, has been developed and diversified through several innovative methods. 
\textit{DeCA}~\cite{a9zhang2023denoising} notes that various models generally produce similar predictions when dealing with clean examples, which reflect real user preferences, whereas predictions on noisy examples vary more significantly. Motivated by this observation, DeCA minimizes the KL divergence between the user preference distributions parameterized by two recommendation models. It simultaneously maximizes the likelihood of data observation, thereby dynamically adjusting weights based on each data point's assessed noise level.
Analogously, \textit{SGDL}~\cite{a27gao2022self} collects memorized interactions identified during the early training stage and uses them as denoising signals. These interactions guide the later ``noise-sensitive" phases of the model's training in a meta-learning fashion. 
\textit{BOD}~\cite{42wang2023efficient} models recommendation denoising as a bi-level optimization problem. It leverages the denoising gradient information generated from the inner optimization process to guide weight determination in the outer optimization for the training data. \textit{AutoDenoise}~\cite{a8lin2023autodenoise} addresses the challenge of highly dynamic data distributions by employing a deep RL-based framework. It incorporates an instance-denoising policy network that adaptively selects noise-free and predictive data instances.

\subsubsection{Contextual Enrichment Training}
Incorporating contextual data types into RS is a crucial approach for identifying and mitigating the impacts of poisoning attacks, thereby significantly enhancing model robustness. By integrating varied data sources—ranging from explicit feedback to social relations and multimedia content—these models can leverage a broader context to better discern user preferences and detect anomalies.
\textit{DUMN}~\cite{a22bian2021denoising} is an exemplary model that utilizes both explicit and implicit feedback. It enhances the representation of implicit feedback by integrating it with explicit feedback through a preference-aware interactive representation component. This component employs gating mechanisms to fuse long-term and short-term user interests, facilitating a nuanced understanding of the evolution of unbiased user preferences. 
\textit{GDMSR}~\cite{a12quan2023robust} improves recommendation robustness by selectively retaining informative social relations. It introduces a self-correcting curriculum learning module to model the confidence in social relations, thereby refining the social graph used in recommendations, enhancing the overall accuracy and trustworthiness of the recommendation system.

For multimedia-rich environments, \textit{Conde}~\cite{a15liu2021concept} leverages semantic information extracted from micro-videos through a three-phase graph convolution process consisting of warm-up propagation, graph denoising, and preference refinement. This method constructs a heterogeneous tripartite graph that connects users, videos, and associated concepts derived from video captions and comments, enhancing the robustness of recommendations by deeply integrating content and user interaction data.
In a related study, \textit{AiD}~\cite{29yuan2023manipulating} focuses on minimizing visual discrepancies between clean and adversarial images while preserving ranking performance. It employs a denoising training strategy tailored to handle visual data, particularly effective in contexts where image-based content significantly influences user preferences.

In specialized recommendation scenarios, such as multi-behavior or sequential recommendations, unique data types further enhance robustness. \textit{DPT}~\cite{a9zhang2023denoising} introduces multiple auxiliary behavior information to enrich the data context. It features a three-stage learning paradigm with a pattern-enhanced graph encoder that harnesses complex patterns as data-driven prior knowledge, aiding in the learning of informative representations and identification of reliable noise signals.
\textit{LoRec}~\cite{c3zhang2024lorec} leverages the general knowledge produced by LLMs to parse textual data and perform user-level calibration, mitigating the impact of malicious data on the model.
\textit{DPGN}~\cite{a16xu2023dpgn}, targeting sequential recommendations in life service applications, adds time periodicity information to model user intentions more accurately. It employs temporal pooling to capture the most representative information from recent behaviors and temporal encoding to manage time intervals, alongside a temporal graph transformer layer to aggregate this temporal information effectively.

\stitle{Remark.}
Robust training aims to fortify the model during its training phase by enhancing or incorporating additional signals from the model or data. This approach contrasts with targeted screening or filtering and significantly boosts the model's overall resistance to poisoning attacks. However, there are several limitations and challenges associated with these methods. For example, one major issue is the computational overhead. Robust training methods often require additional computational resources due to the inclusion of adversarial components, joint training mechanisms, and complex data enrichment processes. This can lead to increased training times and higher operational costs. Another challenge is balancing trade-offs. Achieving a balance between enhancing robustness and maintaining recommendation accuracy can be difficult. As defenders, it is crucial to navigate these challenges carefully to ensure that the efforts to enhance model robustness do not inadvertently degrade the system's performance. On the other hand, while we have provided a comprehensive taxonomy, it is important to note that some areas remain unexplored. For instance, extending beyond simple data incorporation to involve sophisticated strategies that augment multiple views from the raw data itself can provide additional supervision signals. These signals not only enhance the quality of representations but also bolster the model's resilience to attacks. However, these self-supervised aspects have not been fully investigated and represent potential future research directions.

\section{Evaluation}
As we delve deeper into the intricacies of both poisoning attack and countermeasure strategies within RS, it becomes crucial to quantify and understand their effectiveness. In this section, we will explore some commonly used metrics that are instrumental in evaluating the performance of poisoning attack methods and defense mechanisms. These metrics not only help in assessing how significantly attacks can degrade system performance but also gauge the robustness of defense strategies in mitigating such impacts.

\subsection{Attack Metric}
The attack metrics are categorized into two groups based on the objectives of the attacks: the first group focuses on evaluating the performance of targeted attacks, while the second group assesses the efficacy of untargeted attacks.

\stitle{Targeted Attack Metric.}
This group focuses on evaluating the change of target items before and after poisoning attacks. Key metrics used include Recnum@K, HitRatio@K, NDCG@K, NI, and PS.
\begin{itemize}[leftmargin=*]
\item Recnum means the exposure number of target items at top-K recommendation. The formula of Recnum@K is as follows:
\begin{equation}
    \begin{aligned}
        \text{Recnum@K} = \frac{1}{|\mathcal{I}^{T}|} \sum_{u \in \mathcal{U}} {|\mathcal{I}^{T} \cap \mathcal{I}^{rec}_u|},
    \end{aligned}
\end{equation}
\item Hit Ratio means the exposure rate of target items at top-K recommendation. The formula of Hit Ratio@K is as follows:
\begin{equation}
    \begin{aligned}
        \text{Hit Ratio@K} = \frac{1}{|\mathcal{U}||\mathcal{I}^{T}|} \sum_{u \in \mathcal{U}} {|\mathcal{I}^{T} \cap \mathcal{I}^{rec}_u|},
    \end{aligned}
\end{equation}
where $K$ is the number of items in the recommended list, $\mathcal{I}^{T}$ means the set of target items, and $\mathcal{I}_{u}^{rec}$ means the set of items recommended to user $u$.
\item Normalized Discounted Cumulative Gain (NDCG) enhances the evaluation by considering the ranking of target items within the recommendation list. The metric assigns higher values when target items are positioned higher in the list. The formula of NDCG@K is as follows:
\begin{equation}
    \begin{split}
    \rm{NDCG@K} &= \frac{1}{|\mathcal{U}|} \sum_{u \in \mathcal{U}} \frac{{\rm DCG}_{u}@K}{{\rm IDCG}@K},
    \end{split}
    \label{NDCG}
    \end{equation}
    \vspace{-0.5em}
    \begin{equation}
    \begin{split}
    \rm{DCG}_{u}@K &= \sum_{m = 1}^{|\mathcal{I}^{rec}_{u}|} \frac{{rel_m}}{\log_2(m+1)}, 
    \end{split}
    \end{equation}
    
    \begin{equation}
    \begin{split}
    \rm{IDCG}@K &= \sum_{m=1}^{|\mathcal{I}^{T}¥|} \frac{1}{\log_2(m+1)},
    \end{split}
    \end{equation}
where $rel_{m}$ means the corresponding item of index $m$ in user $u$'s recommendation list belongs to target items. 
\item NI means the increased ranks of the target items. The formula of NI is as follows:
\begin{equation}
    \begin{aligned}
        \text{NI} = \frac{1}{|\mathcal{I}^{T}|} \sum_{i \in \mathcal{I}^{T}} {(\widehat{rank}^{i}_u -rank^{i}_u)},
    \end{aligned}
\end{equation}
where $\widehat{rank}^{i}_u$ and $rank^{i}_u$ means the ranking of item $i$ recommended to user $u$ before and after poisoning attack.
\item Prediction Shift (PS) means the increased predicted rating of the target items. The formula of PS is as follows:
\begin{equation}
    \begin{aligned}
        \text{PS} = \frac{1}{|\mathcal{I}^{T}|} \sum_{i \in \mathcal{I}^{T}} {(\widehat{rating}^{i}_u -rating^{i}_u)},
    \end{aligned}
\end{equation}
where $\widehat{rating}^{i}_u$ and $rating^{i}_u$ means the predicted rating of item $i$ to user $u$ before and after poisoning attack.
\end{itemize}

\stitle{Untargeted Attack Metric.} This group focuses on evaluating the change of overall recommendation performance before and after poisoning attacks. Thus, the key metrics used align with the recommendation metric, including MAE, RMSE, $\text{Precision}_{rec}$, and $\text{Recall}_{rec}$.
\begin{itemize}[leftmargin=*]
\item The Mean Absolute Error (MAE) measures the average magnitude of errors between predicted ratings and actual ratings, without considering the direction of these errors. The formula of MAE is as follows:
\begin{equation}
    \begin{aligned}
        \text{MAE} = \frac{1}{|\mathcal{U}||\mathcal{I}^{T}|} \sum_{u \in \mathcal{U} , i \in \mathcal{I}} {(\overline{r}^{i}_u -r^{i}_u)},
    \end{aligned}
\label{MAE}
\end{equation}
where $\overline{r}^{i}_u$ means the predicted rating of item $i$ by user $u$, and $r^{i}_u$ means the real rating of item $i$ by user $u$.
\item Root mean square error(RMSE) squares the differences before averaging them, thus emphasizing larger errors more than MAE. The formula of RMSE is as follows:
\begin{equation}
\text{RMSE} = \sqrt{\frac{1}{|\mathcal{U}||\mathcal{I}^{T}|} \sum_{u \in \mathcal{U}, i \in \mathcal{I}^{T}} (\overline{r}^{i}_u - r^{i}_u)^2},
\label{RMSE}
\end{equation}
\item $\text{Precision}_{rec}$ represents the precision of the recommendations provided by a system. It measures the proportion of recommended items that are relevant to the user. The formula of Precision is as follows:
\begin{equation}
    \begin{aligned}
        \text{Precision}_{rec} = \frac{1}{|\mathcal{U}|} \sum_{u \in \mathcal{U}} {\frac{|\mathcal{I}_{u} \cap \mathcal{I}^{rec}_u|}{\mathcal{I}^{rec}_u}}.
    \end{aligned}
\label{Precision}
\end{equation}
where $\mathcal{I}_{u}$ represents the actual items interacted by user $u$.
\item $\text{Recall}_{rec}$ is another critical metric used to evaluate the performance of RS. While precision measures the proportion of recommended items that are relevant, recall assesses the proportion of relevant items that were actually recommended by the system. The formula for $\text{Recall}_{rec}$ is as follows:
\begin{equation}
    \begin{aligned}
        \text{Recall}_{rec} = \frac{1}{|\mathcal{U}|} \sum_{u \in \mathcal{U}} {\frac{|\mathcal{I}_{u} \cap \mathcal{I}^{rec}_u|}{\mathcal{I}_{u}}}.
    \end{aligned}
\label{Recall}
\end{equation}
\end{itemize}

\subsection{Defense Metric}
On the defense side, the metrics are divided into two groups: the first set evaluates the effectiveness of poisoning filters, and the second set assesses the performance of robust training strategies. 

\stitle{Poisoning Filter Metric.} This group concentrates on assessing the accuracy of identifying poisoned elements within a system. The key metrics commonly utilized for this purpose are $\text{Precision}_{filter}$, $\text{Recall}_{filter}$, $\text{F1}_{filter}$, and AUC.
\begin{itemize}[leftmargin=*]
\item $\text{Precision}_{filter}$: This metric measures the proportion of correctly identified poisoned elements out of all elements flagged by the filter as poisoned. It is defined as:
\begin{equation}
\label{formula15}
\begin{split}
\text{Precision}_{filter}=\frac{\text { \# true positive }}{\text { \# true positive + \# false positive }},
\end{split}
\end{equation}
where $\text{\# true positive}$ denotes the number of correctly identified malicious elements, $\text{\# false positive}$ denotes the number of misclassified real elements, and  $\text{\# false negative}$ denotes the number of attack profiles that are misclassified.
\item $\text{Recall}_{filter}$: this metric assesses the proportion of actual poisoned elements that are correctly identified by the filter. It is calculated as:
\begin{equation}
\label{formula16}
\begin{split}
\text{Recall}_{filter}=\frac{ \text {\# true positive }}{ \text { \# true positive + \# false negative}},
\end{split}
\end{equation}
\item $\text{F1}_{filter}$: F1 is the harmonic mean of precision and recall, offering a balance between the two by penalizing extreme values. The F1 is calculated as follows:
\begin{equation}
\label{formula17}
\begin{split}
\text{F1}=\frac{2 \times \text{Precision}_{filter}\times \text{Recall}_{filter}}{\text{Precision}_{filter}+\text{Recall}_{filter}},
\end{split}
\end{equation}
\item Area Under the Curve (AUC): It means area under the curve of the Receiver Operating Characteristic (ROC) graph, which is a performance measurement for classification problems at various threshold settings. A higher AUC value indicates better performance.
\end{itemize}
\stitle{Robust Training Metric.} 
This group concentrates on assessing the overall performance of the recommendation system. Consequently, the key metrics employed are akin to those used in evaluating untargeted attack metrics, including MAE, RMSE, $\text{Precision}_{rec}$, and $\text{Recall}_{rec}$, referred to Equations~\ref{MAE},~\ref{RMSE},~\ref{Precision}, and~\ref{Recall}.

\section{Discussion}
In this section, we shed light on the existing limitations of existing research and then identify several future research directions worth exploration. These insights aim to address gaps in the field and guide the development of more effective methods and strategies.

\subsection{Limitations}
\stitle{The Lag in Exploring Vulnerabilities of Recommender Systems.} The pace of research aimed at securing RS noticeably trails behind the rapid development of recommendation technologies. Despite a growing awareness of the need to proactively investigate vulnerabilities, the field of RS is characterized by a swift and continuous evolution. This includes frequent updates to recommendation scenarios, diversifying architectures, and evolving learning objectives, each adding layers of new features and complexities. Consequently, these dynamics pose a continuous challenge to the research community, which struggles not only to anticipate potential vulnerabilities but also to develop timely and effective defenses. This lag in defensive readiness leaves systems perpetually at risk of exploitation, highlighting an urgent need for a more agile and forward-thinking approach to keep pace with technological advancements.

\stitle{Out of Sync in Attack and Defense Research.} In current research, both offensive and defensive strategies commonly revolve around a theoretical model of an adversary. Typically, this hypothetical adversary is posited to possess limited capabilities, a scenario not necessarily reflective of intentional bias by researchers but more indicative of the inherent challenges in security research. The core of this issue lies in the concurrent development of attack and defense strategies, where the newest information on adversaries' evolving tactics and technological advancements is often scarce or inaccessible. This information gap significantly hampers researchers' ability to accurately gauge the current threat landscape, leading to potential underestimations of adversary capabilities. As a result, both attack simulations and defense mechanisms may not fully align with the realities of more sophisticated and capable threats, creating a disconnect that could leave systems vulnerable to unanticipated attack vectors. The need for a dynamic, continuously updated approach to understanding and modeling adversary behaviors is critical, urging the adoption of more adaptive and timely strategies to ensure robustness against not only current but also future threats.

\stitle{The Benchmarking Shortfall in Secure RS.} While the broader field of RS thrives with robust benchmarks and collaborative platforms, secure RS continues to grapple with significant gaps in standardized benchmarking tools. This deficiency hampers the ability to perform consistent, reproducible research across the community, crucial for validating and comparing security enhancements. Although platforms like SDLib\footnote{https://github.com/Coder-Yu/SDLib}, RecAD\footnote{https://github.com/gusye1234/recad}, and ARLib\footnote{https://github.com/CoderWZW/ARLib} have been developed to address these needs, they fall short of providing comprehensive coverage across diverse recommendation scenarios and methodologies. These platforms' limitations complicate the task for researchers attempting to rigorously test and improve secure recommendation algorithms under a variety of complex conditions. The establishment of more inclusive and flexible benchmarking platforms is vital to elevate the experimental rigor and innovation rate in secure RS, ensuring it keeps pace with advancements in general RS.

\subsection{Future Directions}
\stitle{Exploration of Novel Victim Contexts.}
The emergence of innovative technologies has given rise to a multitude of recommendation scenarios, consequently expanding the landscape of potential threats. Innovations such as multi-modal recommendation \cite{78yuan2023go}, self-supervised learning-based recommendation \cite{41yu2023self}, and LLM-based recommendation \cite{77wu2023survey} are redefining the complexity and capabilities of modern recommender systems. While these technologies offer significant advancements in personalization and accuracy, they also introduce new layers of vulnerability. Research into the security implications of these emerging technologies is still nascent, barely scratching the surface of potential exploit scenarios. Each new recommendation paradigm brings with it unique challenges and security loopholes that differ significantly from those encountered in more traditional RS setups. As we move forward, the focus must be on understanding these novel contexts in depth—identifying their specific vulnerabilities, assessing the risk they pose, and developing robust defensive strategies that can safeguard against both known and unforeseen threats.

\stitle{Investigation of Sophisticated Malicious Intents.} Current research has identified certain attack intents, but real-world motivations for attacks are likely to be more diverse. One emerging concern is cross-platform manipulation, where attackers might exploit interconnected data ecosystems to influence recommendations on one platform by manipulating another. This type of strategy is particularly potent in environments where user data are extensively shared or linked across different platforms.
Moreover, cultural attacks represent a sophisticated frontier of adversarial strategy. These attacks exploit linguistic or cultural nuances within multilingual RS to subtly skew recommendation outcomes, potentially swaying user perceptions or promoting specific cultural narratives.
Given the expansive variety of potential real-world attacks, a deeper and more comprehensive understanding of these diverse motivations is imperative.

\stitle{Enriching Theoretical Foundation.} While numerous poisoning attack methods have revealed the vulnerabilities of RS, there is a shortage of comprehensive studies on the fundamental principles of poisoning attacks. For instance, a critical aspect is the economic balance: the amount of malicious data injected is constrained by economic factors, and the attack's effectiveness often aims for economic gains. Although the relationship between the attack's cost and its economic impact is acknowledged, its precise nature and dynamics remain unclear. The field of poisoning attacks requires robust theoretical foundations to address such aspects, which would enable more efficient research and reduce reliance on labor-intensive trial-and-error methods.

\stitle{Mixture and Long-Term Impact of Vulnerability.} Current research on poisoning attacks primarily focuses on their immediate effects, such as the rapid injection of malicious data based on the existing data and model configuration. However, this approach may not adequately capture the evolving nature of RS. As RS accumulate more data and undergo continuous model updates, the initial impact of injected poisoning data may diminish or become diluted over time. Therefore, it is crucial to investigate not only the short-term effects of a poisoning attack but also to track and analyze its long-term impacts. Understanding how poisoning data interacts with and influences RS over extended periods can provide deeper insights into the sustained vulnerabilities and resilience of these systems.

\stitle{Data Neutralization for Defense.} An intriguing concept under investigation is the injection of counteractive data into recommendation systems RS to mitigate the negative effects of poisoning data. This strategy, while proposed in other domains~\cite{84chan2019poison}, has been relatively unexplored in the context of poisoning attacks. The primary challenge in implementing this technique lies in achieving this neutralization effectively and efficiently, without imposing additional burdens on the system or compromising the user experience. The exploration of efficient neutralization techniques promises to be a pivotal step in enhancing the resilience of RS against sophisticated and ever-evolving poisoning strategies.

\stitle{Adaptive Enhancement Learning of Attack and Defense Strategies.} Traditionally, researchers exploring attack and defense mechanisms independently might lead to an underestimation of the adversary's capabilities. However, simultaneously researching both aspects can effectively mirror the dynamic nature of real-world cybersecurity confrontations, where the enhancement of one side's tactics inevitably leads to improvements in the other's capabilities. This concept is currently embodied in adversarial learning, which simulates this reciprocal growth in capabilities. While adversarial learning is an existing practice that reflects this idea, there is still room to explore and develop new methods of adaptive learning.

\section{Conclusion}
In recent years, there has been a growing interest in research on secure RS. This survey paper provides a comprehensive review of the current state-of-the-art in the field. It introduces a taxonomy of poisoning attacks, categorizes existing methods into four types, and outlines a defense taxonomy divided into two categories. The summary from both the attack and defense perspectives, enables newcomers to quickly familiarize themselves with the field of secure RS. Finally, we outline future research directions to address the limitations of current studies. We hope it can provide valuable guidance and insights for individuals who are interested in staying up-to-date with the latest developments in secure RS.


{\small
\bibliographystyle{named}
\bibliography{ijcai23}
}

\end{document}